\documentclass[a4paper, amsfonts, amssymb, amsmath, reprint, showkeys,nofootinbib,twoside,unsortedaddress,superscriptaddress,floatfix]{revtex4-1} 

\usepackage[english]{babel}
\usepackage[utf8]{inputenc}
\usepackage[colorinlistoftodos, color=green!40, prependcaption]{todonotes}
\usepackage{aas_macros}

\usepackage{amsthm}
\usepackage{mathtools}
\usepackage{physics}
\usepackage{xcolor}
\usepackage{graphicx}
\usepackage[left=23mm,right=13mm,top=35mm,columnsep=15pt]{geometry}
\usepackage{adjustbox}
\usepackage{placeins}
\usepackage[T1]{fontenc}
\usepackage{lipsum}
\usepackage{csquotes}
\usepackage{booktabs}
\usepackage{array}

\newcolumntype{M}[1]{>{\centering\arraybackslash}m{#1}}

\newcommand{\Lik}{\mathcal{L}}
\newcommand{\TS}{\mathrm{TS}}
\newcommand{\Fermi}{\textit{Fermi}}
\usepackage{threeparttable}
\usepackage{float}
\usepackage[pdftex, pdftitle={Article}, pdfauthor={Author}]{hyperref} 
\begin{document}
\title{Constraints on axionlike particles from a combined analysis of three flaring \textit{Fermi} flat-spectrum radio quasars}

\author{James Davies}
\email[Email address: ]{james.davies2@physics.ox.ac.uk}
\affiliation{University of Oxford, Department of Physics, Oxford OX1 3PJ, United Kingdom}

\author{Manuel Meyer}
\email[Email address: ]{manuel.meyer@desy.de}
\thanks{Also at CP3-Origins, University of Southern Denmark,
Campusvej 55, 5230 Odense M, Denmark.}
\affiliation{Institute for Experimental Physics, University of Hamburg, Luruper Chaussee 149, D-22761 Hamburg, Germany}

\author{Garret Cotter}
\email[Email address: ]{garret.cotter@physics.ox.ac.uk}
\affiliation{University of Oxford, Department of Physics, Oxford OX1 3PJ, United Kingdom}

\date{\today} 

\begin{abstract}
Many theories beyond the Standard Model of particle physics predict the existence of axionlike particles (ALPs) that mix with photons in the presence of a magnetic field. Searching for the effects of ALP-photon mixing in gamma-ray observations of blazars has provided some of the strongest constraints on ALP parameter space so far. Previously, only individual sources have been analyzed. We perform a combined analysis on \textit{Fermi} Large Area Telescope data of three bright, flaring flat-spectrum radio quasars, with the blazar jets themselves as the dominant mixing region. For the first time, we include a full treatment of photon-photon dispersion within the jet, and account for the uncertainty in our $B$-field model by leaving the field strength free in the fitting. Overall, we find no evidence for ALPs but are able to exclude the ALP  parameters $5 \mathrm{ neV}\lesssim m_a\lesssim200$ neV and $g_{a\gamma}\gtrsim 5 \times 10^{-12}$ GeV$^{-1}$ with 95\% confidence.
\end{abstract}


\maketitle

\section{Introduction} \label{sec:intro}

Axions are very light pseudoscalar particles beyond the Standard Model (SM) of particle physics \cite{Peccei_Quinn_1977,Weinberg_78,Wilczek_78}, which provide a theoretical solution to the strong \textit{CP} problem \cite{Peccei_06}. Importantly, axions would couple to photons in the presence of an external magnetic field, with a coupling $g_{a\gamma}$, proportional to its mass $m_a$ \cite{Nakamura_2010,Chadha-Day_Ellis_Marsh_2022}. This coupling would lead to oscillations between photons and axions, comparable to those between neutrino states---an effect that has been the basis for many experimental axion searches (e.g., \cite{Semertzidis_2022}). So far, none have been found. \par
Axionlike particles (ALPs) are similar particles in which the $m_a$/$g_{a\gamma}$ relation is relaxed. Such particles commonly arise in string theories, or as pseudo-Nambu-Goldstone bosons in other SM extensions \cite{Turok_1996,Jaeckel_Ringwald_2010,Ringwald_2014,Irastorza_2018}. ALPs would no longer necessarily solve the strong \textit{CP} problem, but they are good candidates to make up all or some of the dark matter content of the Universe \cite{Preskill_Wise_Wilczek_1983,Abbott_Sikivie_1983,Dine_Fischler_1983,Arias_2012}. This makes them interesting targets for direct and indirect searches too (e.g, \cite{Graham_2015, Isern_2008, Semertzidis_2022}). In particular, ALP-photon mixing in the various magnetic fields found in space could affect observations of astrophysical sources (e.g., \cite{Payez_2015, Lai_2006}). X- and gamma-ray observations of blazars have been used to set some of the strongest constraints on ALP parameter space so far for masses $m_a \lesssim 100$ neV \cite{HESS_2013,Fermi_2016,Reynolds_2020,Buehler_2020,Matthews_2022,Sisk-Reynes_2022}.\par
Blazars are active galactic nuclei (AGN) producing jets of relativistic plasma, which are pointed towards us (within a few degrees). This means their emission is strongly enhanced by relativistic effects; blazars make up some of the brightest gamma-ray sources in the sky \cite{Fermi_4FGL_2020}, though they emit across the entire electromagnetic spectrum, from radio to gamma rays. While the detailed emission mechanisms of blazars are still unclear, the low energy emission is usually considered to be synchrotron photons emitted by electrons in the plasma. The high energy emission is then thought to be inverse-Compton (IC) emission from these same electrons up-scattering either their own synchrotron photons (synchrotron self-Compton), or other background photon fields (external Compton) \cite{Blandford_Rev_2019}. Hadronic models are also possible, for the high energy peak in particular, (e.g., \cite{Petropoulou_2012, Mucke_2003}), though these models may require super-Eddington jets \cite{Zdziarski_2015}. Significantly for ALP searches, a smooth nonthermal distribution of electrons (as produced by, e.g., shock acceleration \cite{Fermi_1949,Marscher_2014}) would produce intrinsically smooth gamma-ray spectra. The presence of ALPs, however, could produce oscillatory spectral features, as the ALP-photon oscillation length could be energy-dependent for some astrophysical environments along the line of sight to the source \cite{deAngelis_Galanti_Roncadelli_Rev_2011}. Looking for these irregularities in individual blazar spectra (NGC 1275 and PKS 2155-304), using their magnetized cluster environments as the mixing region, has been the basis for constraints with \textit{Fermi} Large Area Telescope (LAT), High Energy Stereoscopic System, and Chandra observations \cite{HESS_2013,Fermi_2016,Reynolds_2020}. These searches require good statistics in the gamma-ray data, which is why bright blazars make good targets---especially when in a flaring state. \par
Here, we perform a similar search with a combined analysis of \Fermi-LAT data for three bright, flaring flat-spectrum radio quasars (FSRQs; 3C454.3, CTA 102, and 3C279), using the blazar jets themselves as the main mixing regions. It has been suggested that the strong field in the jet could lead to ALP-photon mixing at higher masses than previously probed by gamma-ray searches \cite{Hochmuth_2007,Bassan_Roncadelli_2009, Sanchez-Conde_2009,Fairbairn_Troitsky_2011,Harris_2014,Tavecchio_2015,Davies_2021}, though so far no search has been performed using it as a mixing region. By combining observations from multiple sources, it should also be possible to strengthen the constraints---within the parameter space probed by all the sources, if an ALP signature is seen in one source, it should be seen in the others too.\par
In Sec. \ref{sec:data} we outline the data selection and spectral analysis performed on the three sources. Then, in Sec. \ref{sec:oscs} we describe the jet and photon-field modeling required to calculate the ALP-photon oscillations produced within the sources. In Sec. \ref{sec:stat}, we then discuss the fitting and statistical analysis used to compare the ALP and no-ALP hypotheses and place limits on ALP parameter space, before presenting the results of our analysis in Sec. \ref{sec:results}. Details of the field structure parameters and the spectral energy distribution (SED) modeling are discussed further in Appendices \ref{sec:appendix_params} and \ref{sec:appendix_seds}, respectively, and the effects of systematics are discussed in Appendix \ref{sec:appendix_sys}.
\section{Data Analysis} \label{sec:data}
The LAT is a pair-conversion, imaging gamma-ray detector on board the \textit{Fermi Gamma-ray Space Telescope} (\Fermi), which measures gamma rays from $30$ MeV up to $>300$ GeV energies \cite{Fermi_fermi_2009}. Our aim is to look for oscillations in \Fermi-LAT gamma-ray spectra caused by ALP-photon mixing. We target the three sources with the brightest flares over the \textit{Fermi} lifetime: 3C454.3, CTA 102, and 3C279 \cite{Meyer_Scargle_Blandford_2019}. We use \texttt{FERMIPY} \texttt{v1.0.1}\footnote{\url{https://fermipy.readthedocs.io} as accessed on Oct 5, 2022} \cite{Wood_fermipy_2017} and \textit{Fermi Science Tools} \texttt{v2.0.8}\footnote{\url{https://fermi.gsfc.nasa.gov/ssc/data/analysis/software/} as accessed on Oct 5, 2022} for the analysis.

\subsection{Data selection}\label{subsec:data_selection}
Initially, we analyze each source over a significant fraction of the \textit{Fermi}-LAT lifetime ($11.7$ years between August 4, 2008 and April 1, 2020) to get an average model for each region of interest (ROI). We choose an energy range of 100 MeV to 500 GeV. This long-term ROI model can then be used as an initial condition for fitting the flare observations. Each ROI is centered on the respective source and has a size of $15^\circ\times15^\circ$. To avoid including gamma-rays produced from the Earth limb, we only use events with a zenith angle $\theta_z\leq90^\circ$. We choose a spatial binning of $0.1^\circ$ pixel$^{-1}$. We use the \texttt{\textbf{P8R3\_SOURCE\_V2}} instrument response functions\footnote{See: \url{https://fermi.gsfc.nasa.gov/ssc/data/analysis/documentation/Cicerone/Cicerone_LAT_IRFs/IRF_overview.html} as accessed on Oct 5, 2022} (IRFs) and only use events that pass the \texttt{\textbf{P8R3 SOURCE}} event selection. Because we are looking for spectral oscillations, we make use of the \texttt{EDISP} event classes available with the \texttt{\textbf{Pass 8}} IRFs \cite{Atwood_2013}. Events are classified into four classes, \texttt{EDISP0} to \texttt{EDISP3}, depending on the quality of their energy reconstruction (worst to best respectively). These classes each contain a similar number of events and are analyzed separately with their corresponding IRFs. This allows us to extract the best spectral information from the data possible.
  For the long-term analysis, we use eight energy bins per decade. Then, for the flare analyzes, we choose the binning so as to reach the smallest resolvable energy scale. This is done by extracting the detector response matrices for our observations and choosing the bin width to match the minimum $\Delta E/E$ for the best energy dispersion class (\texttt{EDISP3}). This gives 65 (3C454.3), 67 (CTA 102), and 61 (3C279) bins per decade for our sources.

\subsection{ROI fitting}\label{subsec:data_fitting}
\begin{figure}[H]
  \centering
    \includegraphics[width=0.48\textwidth]{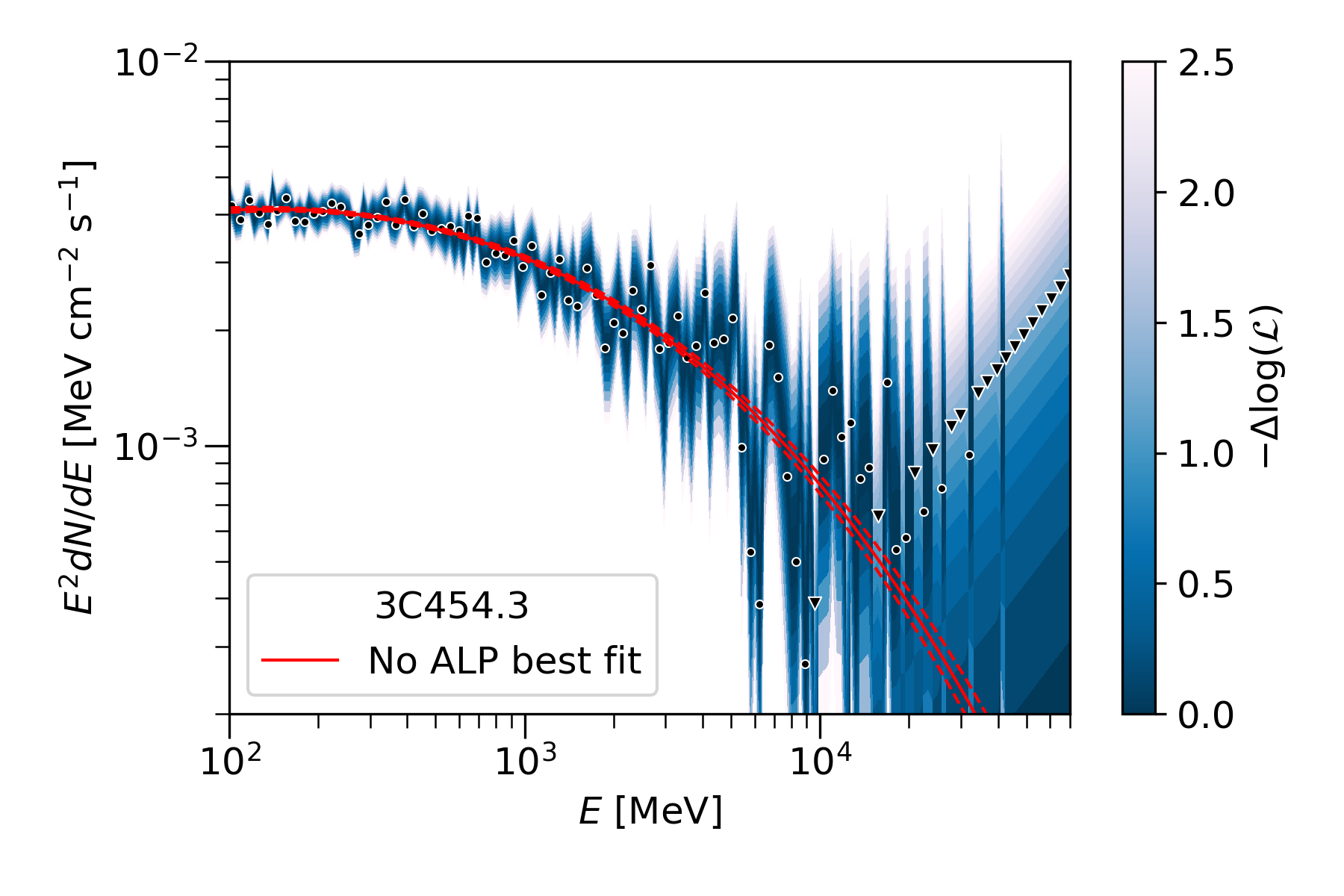}
    \includegraphics[width=0.48\textwidth]{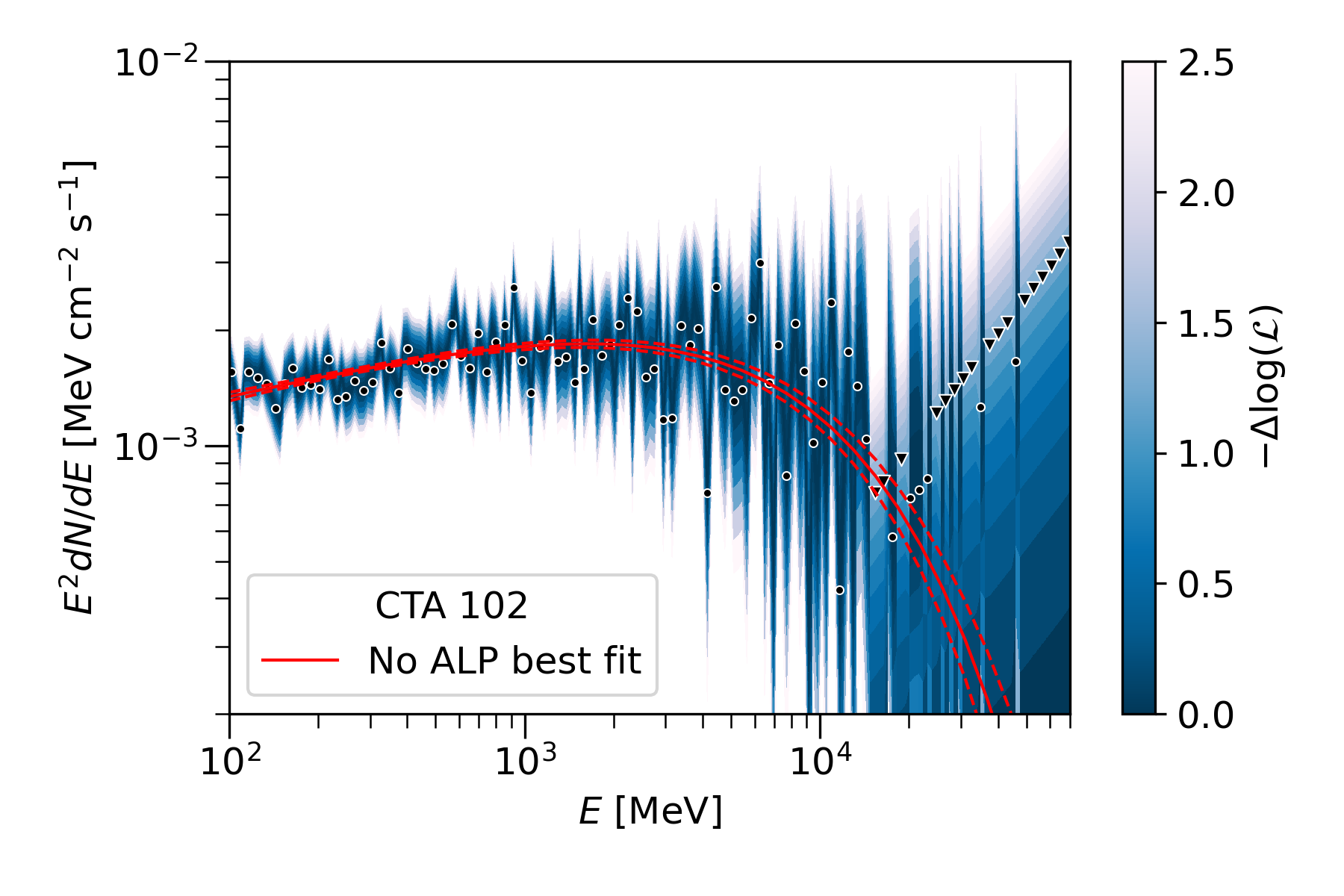}
    \includegraphics[width=0.48\textwidth]{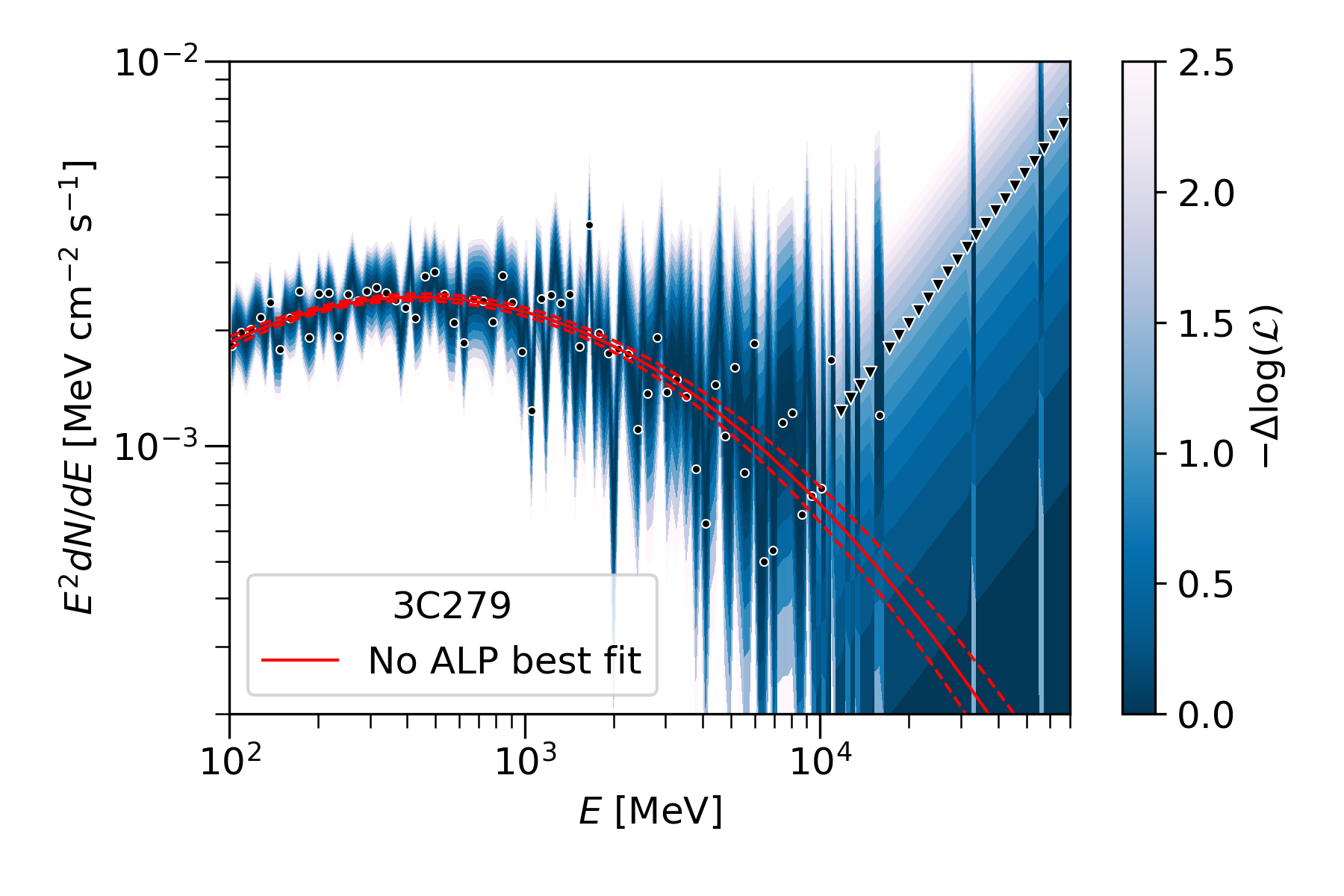}
  \caption{SEDs for our sources during the flares: 3C454.3 (top), CTA 102 (middle), 3C279 (bottom). Best-fit spectral parameters (corresponding to the red lines) and time ranges used are listed in Table \ref{tab:flares}. Black points show detections and triangles show 95\% upper limits. Shaded regions show the likelihood curves for each energy bin. For clarity, only points for every other energy bin are plotted.}
  \label{fig:seds}
\end{figure}

First, we optimize the ROI model for each of our sources, for all the event types combined, over the entire 11.7 year time range defined above. The initial model includes every point source in the 4FGL catalogue (Data Release 1) \cite{Fermi_4FGL_2020} and the standard diffuse isotropic and galactic background templates\footnote{We use \texttt{iso\_P8R3\_SOURCE\_V2} templates for the isotropic background for each \texttt{EDISP} class, and \texttt{gll\_iem\_v07.fits} for the galactic background, which can be found here: \url{https://fermi.gsfc.nasa.gov/ssc/data/access/lat/BackgroundModels.html} as accessed on Oct 5, 2022}. We then free the normalization of all sources, including the diffuse backgrounds. Point sources within $5^\circ$ of the ROI center or with test statistic $\TS>10$ have the rest of their spectral parameters freed too ($\TS$ is the log-likelihood ratio of the likelihoods with and without the source). We free the spectral index of the galactic background as well. These free model parameters are then fitted to the data. Within this fitted ROI, we search for new point sources to add to the model by calculating a $\TS$ map. This is done by adding a potential point source (with a power-law index, $\Gamma=2$) at each pixel of the ROI and calculating its $\TS$. Sources with $\sqrt{\TS}\geq5$ are then added to the overall ROI model at the position which gives the highest $\TS$. We then reoptimize the entire ROI, and repeat the process until no more sources are found; overall we find four new sources each for 3C454.3 and 3C279, and three for CTA 102. This gives the final best-fit model for each of our ROIs over the long-term time period. The time ranges used for our flares are taken from the light-curve analysis of \cite{Meyer_Scargle_Blandford_2019} and are listed in Table \ref{tab:flares}. For each of these ranges, we redo the above analysis using these final best-fit ROI models, including the new sources, as the initial conditions---this time only freeing the galactic background and sources that still have $\TS>10$. We fit each event type separately, treating them as separate measurements (as in \cite{Fermi_2016}). Once fitted ROI models have been found for each of our flares, we calculate SEDs for our sources of interest. Following the 4FGL catalogue, the spectra of 3C454.3 and CTA 102 are both best fitted by a power law with a superexponential cutoff:
\begin{equation}\label{eq:plec}
    \frac{dN}{dE} = N_0\left(\frac{E}{E_0}\right)^{-\Gamma_1}\exp\left\{-\left(\frac{E}{E_c}\right)^{\Gamma_2}\right\},
\end{equation}
whereas the 3C279 spectrum is best fitted with a log-parabola,
\begin{equation}\label{eq:logpar}
    \frac{dN}{dE} = N_0 \left(\frac{E}{E_0}\right)^{-\left(\Gamma_1 + \kappa \ln(E/E_0)\right)}.
\end{equation}
$N$ is the number of photons received per unit area per unit time at photon energy $E$, $N_0$ is the spectral normalization, $E_0$ is the reference energy, $E_c$ is the cutoff energy, and $\Gamma_1$, $\Gamma_2$ and $\kappa$ are indices. Each event type will have different best-fit spectral parameters; those for a combined event-type analysis are shown in Table \ref{tab:flares}, and the corresponding SEDs ($E^2 dN/dE$) are shown in Fig. \ref{fig:seds}. For clarity, only every other energy bin is plotted, but we utilize the full energy resolution in our analysis steps.\par
\begin{table*}
 \small
 \caption{Time ranges ($t_\mathrm{start}$ to $t_\mathrm{end}$) and best-fit spectral parameters (see Eqs. \ref{eq:plec} and \ref{eq:logpar}) for a combined event-type analysis of the flares. $N_0$ is the spectral normalization, $E_0$ is the reference energy, $E_c$ is the cutoff energy, and $\Gamma_1$, $\Gamma_2$ and $\kappa$ are indices.}
  \centering
  \begin{tabular}{M{0.08\textwidth} M{0.08\textwidth} M{0.2\textwidth} M{0.13\textwidth} M{0.13\textwidth} M{0.13\textwidth} M{0.13\textwidth} M{0.06\textwidth}}
    \toprule\toprule
    $t_\mathrm{start}$ & $t_\mathrm{end}$ & $N_0$ & $\Gamma_1$ & $\kappa$ & $E_c$ & $\Gamma_2$ & $E_0$ \\ \\
    MJD & MJD & [$10^{-9}$ MeV$^{-1}$ cm$^{-2}$ s$^{-1}$] & & & MeV & & MeV \\
    \midrule
    \multicolumn{8}{c}{3C454.3 (4FGL J2253.9+1609)} \\ \\
    55,516.55 & 55,525.48 & $525.7\pm48.46$ & $1.443\pm0.029$ &  & $2.614\pm0.414$ & $0.227\pm0.0046$ & $410.0$ \\
    \midrule
    \multicolumn{8}{c}{CTA 102 (4FGL J2232.6+1143)} \\ \\
    57,749.10 & 57,754.09 & $2.113\pm0.175$ & $1.813\pm0.036$ &  & $9848\pm2315$ & $0.819\pm0.134$ & 1000 \\
    \midrule
    \multicolumn{8}{c}{3C279 (4FGL J1256.1-0547)} \\ \\
    57,188.07 & 57,189.94 & $12.47\pm0.262$ & $2.004\pm0.018$ & $0.126\pm0.013$ &  &  & 442.1 \\
    \bottomrule\bottomrule
  \end{tabular}
  \label{tab:flares}
\end{table*}

For each event type $k$, we then extract likelihood curves\footnote{Throughout, we use the shorthand $\Lik(\mu)\equiv\Lik(\mu|x)$, where $x$ is the observed data.} $\Lik^k(\mu_i)$, in each energy bin $i$, as a function of expected counts $\mu_i$, from these best-fit SEDs\footnote{We extract bin-by-bin likelihood curves using the SED function within \texttt{FERMIPY}. This function changes the normalization in each energy bin and recomputes the likelihood at each point, taking energy dispersion into account.} (shown as blue bands in Fig. \ref{fig:seds}). As can be seen, the best statistics are at low energies, and no detected emission is seen at energies above about 80 GeV. These curves can then be used to perform a likelihood ratio test between models with and without ALPs (see Sec. \ref{sec:stat}). For each event type, the total likelihood for the no-ALPs model is then
\begin{equation}\label{eq:L_0}
    \Lik^k_{0} = \prod_i \Lik^k(\Bar{\mu}_i)
\end{equation}
for each source, where $\Bar{\mu}_i$ are the expected counts from the best-fit spectral models, including all photon absorption (see Fig. \ref{fig:abs} below), but without ALPs (best-fits with ALPs will later be denoted with a hat as opposed to a bar).

\section{ALP-photon oscillations} \label{sec:oscs}
To test whether an ALP signature is present in the data, we must model the spectral oscillations caused by ALP-photon mixing.\par
In general, a photon of energy $E$ propagating in a homogeneous field $B$, (with a component $B_T$, transverse to the photon direction of motion, and parallel to one of the photon polarization states) will oscillate into an ALP with mass $m_a$ and coupling $g_{a\gamma}$, with a wave number \cite{Raffelt_1988,deAngelis_Galanti_Roncadelli_Rev_2011,Dobrynina_2015}
\begin{equation}\label{eqn:delta_osc}
\begin{split}
    \Delta_{osc} = \Bigg[&\left\{\frac{m_a^2 - m_T^2}{2E} + E\left(b + \chi + i\frac{\Gamma_{\gamma\gamma}}{2E}\right)\right\}^2 \\
    &+ (g_{a\gamma}B_T)^2\Bigg]^{\frac{1}{2}},
\end{split}
\end{equation}
where $m_T$ is the effective mass of the photon (see Ref. \cite{Davies_2021} for the calculation within the jet). $\chi$ and $\Gamma_{\gamma\gamma}$ are the total dispersion and absorption terms for the surrounding photon fields respectively, and
\begin{equation}\label{eqn:b}
    b=\frac{7\alpha}{90\pi}\left(\frac{B_T}{B_\mathrm{cr}}\right)^2
\end{equation}
is the vacuum QED term describing dispersion off the magnetic field, with $B_\mathrm{cr}$ the critical magnetic field $B_{cr}=m_e^2/|e|\sim4.4\times10^{13}$ G, where $e$ is the electric charge. Assuming absorption is small, this means there are two so-called "critical energies", around which the oscillation length depends strongly on energy (and so ALP-photon mixing could lead to oscillations in energy spectra):
\begin{equation}\label{eq:e_low}
    E_\mathrm{crit}^\mathrm{low} = \frac{|m_a^2 - m_T^2|}{2g_{a\gamma}B_T}
\end{equation}
which depends on the effective mass difference between the ALP and the photon, and
\begin{equation}\label{eq:e_high}
    E_\mathrm{crit}^\mathrm{high} = \frac{g_{a\gamma}B_T}{b+\chi},
\end{equation}
which depends on the dispersion terms. For astrophysical plasma environments, these energies can be in the gamma-ray energy range for interesting ALP parameters. This has been the basis of previous searches, and is the basis of ours.\par
In order to model these spectral oscillations, then, we need a model of the field strength and orientation of the magnetic fields along the sight between us and the gamma-ray sources. This allows us to calculate the photon survival probability, $P_{\gamma\gamma}$, i.e., the probability that the emitted photon arrives as a photon at Earth as a function of photon energy (taking into account both photon-ALP conversion and absorption via pair production). The magnetic fields we include along the line of sight are the jet field and the galactic magnetic field (GMF) of the Milky Way, as we choose a mass range where the intergalactic magnetic field does not contribute strongly (see Sec. \ref{sec:stat} below) and these sources are not thought to be in highly magnetized clusters. For the GMF, we use the model of Ref. \cite{Jansson_Farrar_GMF_2012}, as used in, e.g. \cite{Fermi_2016,CTA_Gpropa_2021}. We also include extragalactic background light (EBL) absorption for propagation through intergalactic space, using the model of Ref. \cite{Dominguez_EBL_2011}. The dominant mixing region we are using, however, is the jet field. We use the \texttt{gammaALPs} \texttt{PYTHON} package\footnote{Hosted on GitHub (\url{https://github.com/me-manu/gammaALPs}) and archived on Zenodo \cite{gammaALPs_2021}. Data files and an example notebook connected to this publication are also available at \cite{zenodo_data_2023}.} to solve the ALP-photon mixing equations throughout---see \cite{Meyer_ICRC_2021} for an overview.

\subsection{Jet modeling} \label{subsec:jetmodel}
 For mixing within a jet, the detailed structure of the jet field needs to be taken into account, as well as dispersion and absorption from the various photon fields within the jet \cite{Tavecchio_2015, Davies_2021,Davies_2022}.\par
We use the Potter \& Cotter jet framework (PC, see \cite{Potter_Cotter_NC_2015}) for the overall jet properties (shape of the field strength, bulk Lorentz factors, and electron density)\footnote{Our jet model is available within the \texttt{gammaALPs} package.}, as discussed in the context of ALP-photon mixing in \cite{Davies_2021}. The structure of the PC jet model is a parabolic, magnetically dominated accelerating jet base, which transitions to a decelerating ballistic conical jet in rough energy equipartition at $r_\mathrm{tr}\sim 10^5 r_g$ from the black hole, where $r_g$ is the gravitational radius, which depends on the black hole mass as $r_g=2GM/c^2$. In the accelerating region ($r\leq r_\mathrm{tr}$) the bulk Lorentz factor is $\Gamma\propto r^{1/2}$, and it is $\Gamma \propto \log(r)$ in the decelerating region ($r_\mathrm{tr}<r\leq r_\mathrm{jet}$), where $r_\mathrm{jet}$ is the jet length. This leptonic jet framework is consistent with theory, observation, and simulations, and is capable of reproducing broadband steady-state SEDs for many blazars \cite{Potter_Cotter_NC_2015}.\par
For the location of the gamma-ray emitting regions during the flares $r_{em}$, we use the lower limits found in Ref. \cite{Meyer_Scargle_Blandford_2019}, derived from the absence of attenuation due to pair production with broad line region (BLR) photons in the gamma-ray spectra.
We use the $B(1 \text{pc})$ values found in \cite{Potter_Cotter_NC_2015} from fits with the PC model to set the initial value of the magnetic field strength $B_0$, which is then left free in the fitting (see Sec. \ref{sec:stat} below). These initial values are slightly lower than those derived from very-long-baseline interferometry core-shift measurements for each of our sources in Ref. \cite{Zamaninasab_2014}, where they assume a conical jet throughout\footnote{We also found that these larger values were incompatible with the SED modeling performed in Appendix \ref{sec:appendix_seds}}.
The electron density varies as $n_e\propto R^{-2}$, where $R$ is the jet width, with the value at $r_\mathrm{tr}$ derived from energetic equipartition. Values for the jet parameters used are listed in Table \ref{tab:jet_props}.\par
\begin{figure}
  \centering
    \includegraphics[width=0.48\textwidth]{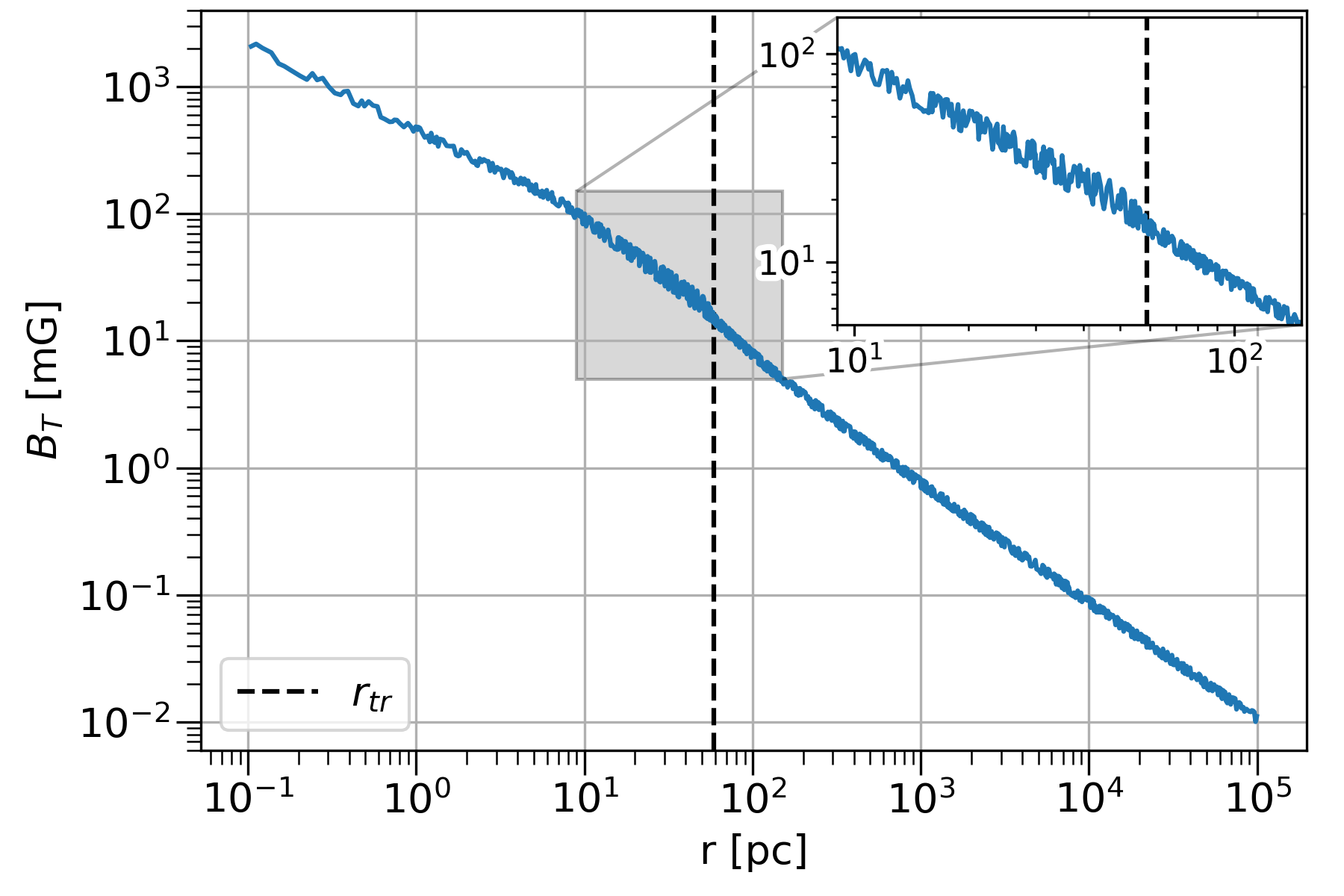}
  \caption{One realization of the transverse component of the magnetic field, $B_T$, for 3C454.3, using the parameters $f=0.3$, $\alpha=1$, and $r_T=r_\mathrm{tr}$. Vertical dashed line shows $r_\mathrm{tr}$.}
  \label{fig:Bs_example}
\end{figure}

\begin{table}
 \caption{Jet properties for our sources: $r_\mathrm{em}$ and $r_\mathrm{tr}$ are the locations of the emission region and jet-base transition region respectively; $r_\mathrm{jet}$ is the jet length; $B_0$ is the field strength, $n_e$ is the electron density; $\Gamma$ is the bulk Lorentz factor; and $r_T$, $\alpha$, and $f$ are the field structure parameters.}
  \centering
  \begin{tabular}{m{0.1\textwidth}m{0.075\textwidth} m{0.09\textwidth}m{0.09\textwidth}m{0.065\textwidth}}
    \toprule\toprule
    Parameters & Unit & 3C454.3 & CTA 102 & 3C279 \\
    \midrule
    $r_\mathrm{em}$& pc & 0.103 & 0.104 & 0.016 \\
    $r_\mathrm{tr}$ & pc & 59.8 & 56.6 & 47.9 \\
    $r_\mathrm{jet}$& kpc & 100 & 75.3 & 32.4 \\
    $B_0(r_\mathrm{tr})$& mG & 16.0 & 26.2 & 6.28 \\
    $n_e (r_\mathrm{tr})$ & cm$^{-3}$ & 4.7 & 2.5 & 5.0 \\
    $\Gamma(r_\mathrm{tr})$&  & 60 & 52 & 37 \\
    $\Gamma(r_\mathrm{jet})$&  & 35 & 29 & 18 \\
    $r_T$& pc & 59.8 & 56.6 & 47.9 \\
    $\alpha$&  & 1 & 1 & 1 \\
    $f$&  & 0.3 & 0.3 & 0.3 \\
    \bottomrule\bottomrule
  \end{tabular}
  \label{tab:jet_props}
\end{table}
We then model the detailed field structure as in Ref. \cite{Davies_2021}, with a tangled component ($B_t$) and a helical component ($B_h$) that transitions from poloidal to toroidal as $r$ increases down the jet. A constant fraction, $f$, of the total field energy density is in the tangled component:
\begin{equation}
    \frac{B^2_{t}}{B^2_{h}} = \frac{f}{1-f}.
\end{equation}
The radius at which the helical field component becomes toroidal is $r_T$; the transverse component of the helical field varies as $B_T\propto r^{-\alpha}$ for $r<r_T$. The three parameters $f$, $r_T$, and $\alpha$ therefore govern the detailed field structure (along with a treatment of the coherence length of the tangled field).
Ideally, these parameters would be allowed to vary in the fit in the same way $B_0$ is. However, because of computational constraints ($P_{\gamma\gamma}$ has to be recalculated every time one of them changes), it is necessary to fix them. In Appendix \ref{sec:appendix_params}, we motivate our choices for these parameters from observation and simulations, and show that varying them would be unlikely to strongly affect our final results. For all our sources, we use $\alpha=1$ and $r_T = r_\mathrm{tr}$. Figure \ref{fig:Bs_example} shows one example field realization for 3C454.3 with this set of parameters.

\subsection{Photon fields}\label{subsec:photon_fields}
\begin{table*}
 \caption{Parameters used for the AGN fields. From the top: disk luminosity, black hole mass (for setting the gravitational radius $r_g$), inner disk radius, outer disk radius, $H\beta$ line luminosity, $H\beta$ line radius, torus temperature, inner torus radius, outer torus radius, semiminor to semimajor axis ratio of torus cross section.}
  \centering
  \begin{threeparttable}
  \begin{tabular}{M{0.1\textwidth}M{0.1\textwidth} M{0.18\textwidth}M{0.18\textwidth}M{0.18\textwidth}}
    \toprule\toprule
    Parameter & Unit & 3C454.3$^\mathrm{a}$ & CTA 102$^\mathrm{b}$ & 3C279$^\mathrm{c}$ \\
    \midrule
    \multicolumn{5}{c}{Disk} \\ \\
    $L_{disk}$ & erg s$^{-1}$ & $2\times10^{46}$  & $4\times10^{46}$  & $3\times10^{45}$ \\
    $M_{BH}$ & $M_{\odot}$ & $1.2\times10^{9}$ & $8.51\times10^{8}$ & $3\times10^{8}$ \\
    $R_{in}$ & $r_g$ & $6$ & $6$ & $6$ \\
    $R_{out}$ & $r_g$ & $200$ & $200$$^\mathrm{a}$ & $430$ \\
    \midrule
    \multicolumn{5}{c}{Broad Line Region} \\ \\
    $L_{H\beta}$ & erg s$^{-1}$ & $4.18\times10^{43}$ & $4.93\times10^{43}$ &  $1.73\times10^{43}$$^\mathrm{b}$\\
    $R_{H\beta}$ & cm & $4.3 \times 10^{17}$ &  $6.1 \times 10^{17}$ & $2.8 \times 10^{17}$ \\
    \midrule
    \multicolumn{5}{c}{Torus} \\ \\
    $\Theta$ & K &$1000$ &$1000$ &$500$  \\
    $R_1$ & cm & $1.6\times10^{19}$ & $1.6\times10^{19}$$^\mathrm{a}$ & $1.6\times10^{19}$$^\mathrm{a}$  \\
    $R_2$ & cm & $1.6\times10^{20}$ & $1.6\times10^{20}$$^\mathrm{a}$ & $1.6\times10^{20}$$^\mathrm{a}$ \\
    $b/a$$^\mathrm{d}$ &  & 0.527 ($f_c=0.6$) & 0.527 ($f_c=0.6$) & 0.527 ($f_c=0.6$) \\
    \bottomrule\bottomrule
  \end{tabular}
  \begin{tablenotes}
      \small
      \item $^\mathrm{a}$ Values taken from Ref. \cite{Finke_2016}, unless marked otherwise.
      \item$^\mathrm{b}$ From Ref. \cite{Meyer_Scargle_Blandford_2019} unless marked otherwise.
      \item$^\mathrm{c}$ From Ref. \cite{HESS_3C279_1415_2019} unless marked otherwise. For $R_{H\beta}$ for 3C279 we use the relation of $R_{H\beta}$ and $R_{Ly\alpha}$ in Ref. \cite{Finke_2016} to convert from $R_{Ly\alpha}$.
      \item$^\mathrm{d}$ From Ref. \cite{Davies_2022}.
    \end{tablenotes}
  \end{threeparttable}
  \label{tab:fields}
\end{table*}

As well as the magnetic field structure, background photon fields can also affect ALP-photon mixing \cite{Dobrynina_2015,Davies_2022} (see $\chi$ and $\Gamma_{\gamma\gamma}$ in Eqs. \eqref{eqn:delta_osc} and \eqref{eq:e_high}). This is because the oscillations are sensitive to slight differences in propagation between the ALP and photon states. Specifically, gamma rays will be affected by photon-photon dispersion and absorption via pair production from background photon fields, whereas ALPs will not.  The fields we would expect within FSRQs are those from the central AGN (accretion disk, BLR, dust torus), starlight (extragalactic and from the host galaxy), the cosmic microwave background (CMB), and synchrotron photons from the jet plasma itself. Reference \cite{Davies_2022} investigated the effects of all these fields on mixing within 3C454.3. They found that for emission regions on the scale of the AGN fields, dispersion off of them will dominate and should be included in the calculations. In particular, for our $r_{\mathrm{em}}\sim0.1$ pc, we expect the BLR and torus fields to be the most important, as the disk is only relevant at much smaller scales. Dispersion from the CMB can play a large role within the jet at energies above 100 GeV, but, as can be seen from Fig. \ref{fig:seds}, we are only interested in lower energies. (In fact, gamma-rays at these energies would likely be absorbed by BLR photons in our sources anyway, see Fig. \ref{fig:abs} below). modeling of the starlight and synchrotron fields therefore does not have to be extremely precise. Nevertheless, we include all the photon fields for each of our sources, using the same method and models as Ref. \cite{Davies_2022}. The various parameters we use for the AGN fields, along with their sources, are given in Table \ref{tab:fields}.\par
The $\chi$ and $\Gamma$ calculations depend on the geometry as well as the photon energies and energy densities of the background fields. The disk is modeled as flat, extending radially in the plane perpendicular to the jet between $R_{in}$ and $R_{out}$, with each radius between the two emitting at only one energy (as in Ref. \cite{Finke_2016}).
We use $R_{in}=6 r_g$ for all sources, the expected inner disk radius for a Schwarzschild black hole. \par
The BLR is modeled as a series of concentric rings, each corresponding to an emission line, and also emitting at only one energy. The radii and luminosities of the lines can be derived from those of the H$\beta$ line for each source (we use all the lines in the Appendix of \cite{Finke_2016}).\par
We use the torus model described in Ref. \cite{Davies_2022}---an extension of the flat model of Ref. \cite{Finke_2016} to include an elliptical torus cross section. Each torus emits at a single energy, depending on its temperature, $\Theta$, and the fraction of disk radiation reemitted in each case is assumed to be $\xi_{dt}=0.1$ (as in \cite{Finke_2016}). All tori for our sources are given the same size and shape. They extend radially between $R_1$ and $R_2$ and have a height so as to give a covering fraction---the fraction of the sky obscured by the torus from the point of view of the black hole---of $f_c=0.6$, which is considered typical (e.g., \cite{Calderone_2012}). The cloud number density within the torus decreases with $R$ from the black hole $\propto R^{-1}$ for all our sources (see \cite{Finke_2016,Davies_2022} for details). \par
We also include the EBL, starlight, and CMB fields exactly as described in Ref. \cite{Davies_2022}, though, as mentioned above, they are subdominant. In order to model the (also subdominant) synchrotron photon field within the jet, we then follow Ref. \cite{Davies_2022} in modeling broadband SEDs for each of our sources (in both flaring and steady states) and compare them to observations (see Appendix \ref{sec:appendix_seds}). This also enables us to confirm the overall self-consistency of our jet and photon-field models.\par
Figure \ref{fig:abs} shows photon survival probability $P_{\gamma\gamma}$ as a function of observed energy for each of our sources, displaying the total absorption from all the photon fields (including intergalactic EBL absorption).
The absorption rates are calculated as in Ref. \cite{Davies_2022}, and are included in every calculation, both with and without ALPs. We note that, even though $B_0$ changing in the fit would, in principle, change the synchrotron field, we keep all the fields fixed. This is a good approximation because the synchrotron field hardly affects the dispersion, and never affects the absorption (see \cite{Davies_2022} and Appendix \ref{sec:appendix_seds}).

\begin{figure}
  \centering
    \includegraphics[width=0.48\textwidth]{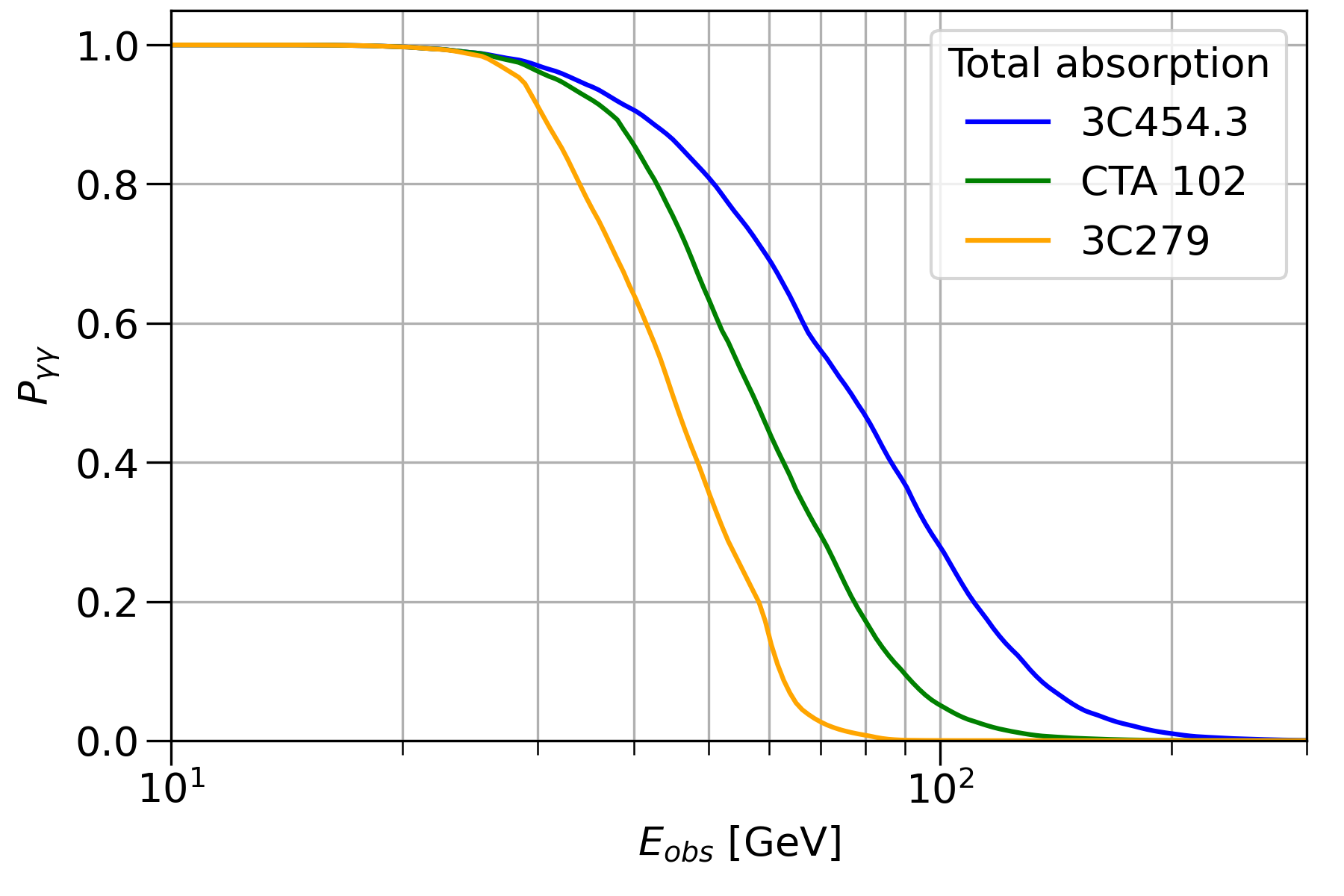}
  \caption{Photon survival probability $P_{\gamma\gamma}$ as a function of observed energy for each of our sources, displaying the total absorption for each of our sources, including all photon fields.}
  \label{fig:abs}
\end{figure}
\section{Statistical methods} \label{sec:stat}

\begin{figure}
  \centering
    \includegraphics[width=0.48\textwidth]{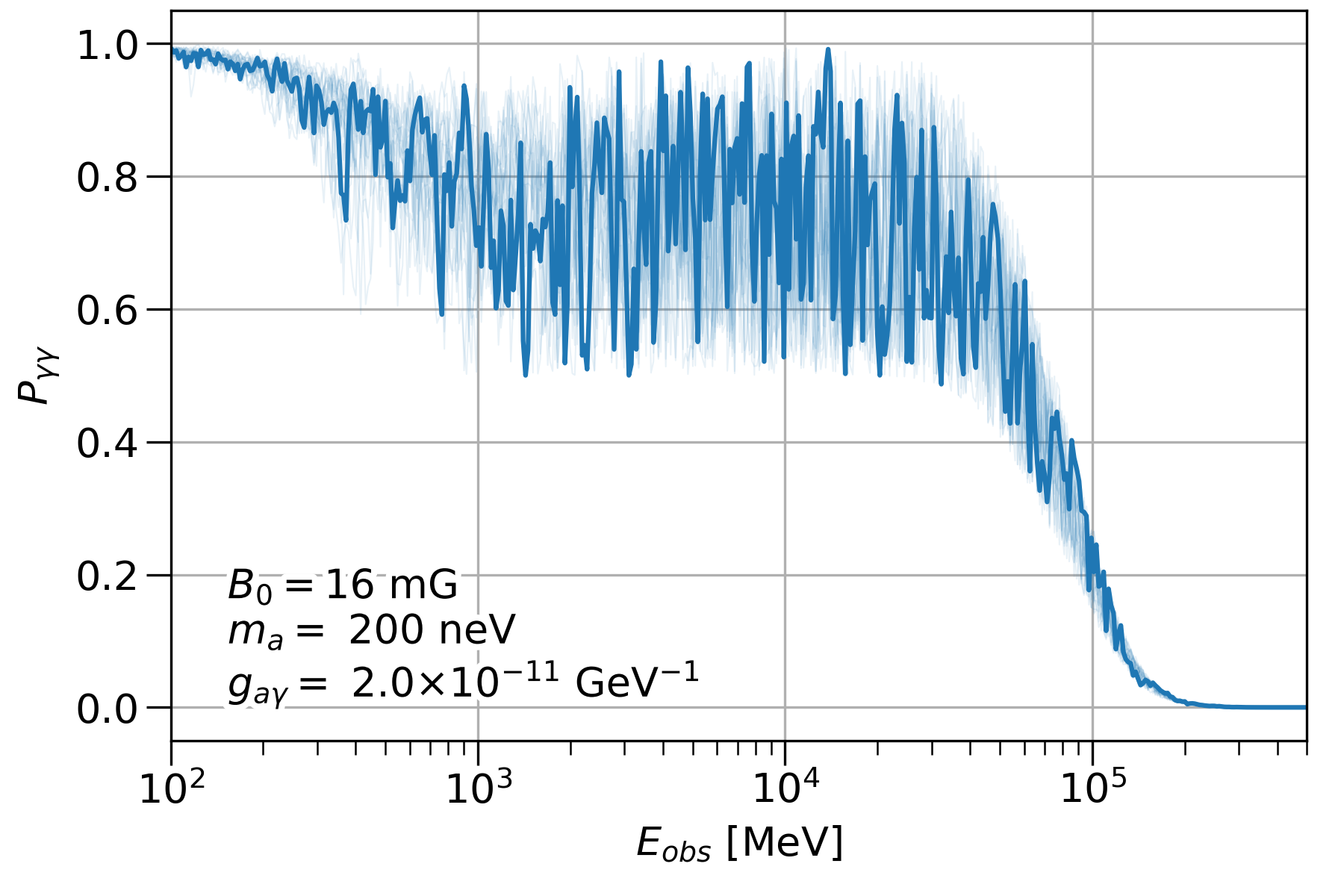}
  \caption{Example photon survival probability $P_{\gamma\gamma}$ for one realization of the 3C454.3 field, using $f=0.3$, $\alpha=1$, and $r_T=r_\mathrm{tr}$. Twenty more realizations are shown in the background. Dispersion and absorption off of the background photon fields are included.}
  \label{fig:pggs_example}
\end{figure}

We are now in a position to compute photon survival probabilities, $P_{\gamma\gamma}(m_a,g_{a\gamma},\mathbf{B}_j)$ for ALP-photon beams propagated through our sources, where $\mathbf{B} = (B_0,f,\alpha,r_T)$ and $j$ denotes a realization of the random magnetic field.
Figure \ref{fig:pggs_example} shows an example $P_{\gamma\gamma}(E_\mathrm{obs})$ for one pair of ALP parameters for 3C454.3, including both dispersion and absorption from the background fields. Our aim is to compare models with ALPs to the observed \textit{Fermi} data. We follow the statistical methods of, e.g., Refs. \cite{HESS_2013,Fermi_2016,CTA_Gpropa_2021} closely. For a random $B$-field realization $j$, and spectral parameters $\boldsymbol{\theta}$, the expected counts including ALPs are then
\begin{equation}\label{eq:mu_alp}
    \mu_i(m_a, g_{a\gamma},j) = \langle P_{\gamma\gamma}(m_a,g_{a\gamma},\mathbf{B}_j) \rangle_i \cdot \mu_i(\boldsymbol{\theta}),
\end{equation}
where $\langle P_{\gamma\gamma}\rangle_i$ denotes the average over energy bin $i$, which is necessary because $P_{\gamma\gamma}$ can vary on energy scales much smaller than the bin width. It is worth noting that possible uncertainties in the instrument response functions used within \texttt{FERMIPY} could possibly slightly affect this expression. In Appendix \ref{sec:appendix_sys}, we show that the inclusion of an extra shift or smear in the energy reconstruction or dispersion would not greatly affect our overall results. We calculate $P_{\gamma\gamma}$ at 500 fine energy bins, logarithmically spaced across our energy range, before averaging. For each source, one set $(m_a,g_{a\gamma},\mathbf{B}_j,\boldsymbol{\theta})$ corresponds to a likelihood,
\begin{equation}\label{eq:L_alp}
    \Lik_{\mathrm{ALP}}(m_a,g_{a\gamma},\mathbf{B}_j,\boldsymbol{\theta}) = p(B_0) \prod_i \Lik(\mu_i(m_a, g_{a\gamma},j)),
\end{equation}
where $\Lik(\mu_i)$ are the likelihood curves extracted from the \textit{Fermi} SEDs, evaluated at the expected ALP counts, and
\begin{equation}\label{eq:B_prior}
    p(B_0) = \exp\left\{-\frac{1}{2}\left(\frac{B_0 - \Bar{B}_0}{\sigma_B}\right)^2\right\},
\end{equation}
is the prior on $B_0$, which takes the form of a Gaussian. $\Bar{B}_0$ is the initial value used (see Table \ref{tab:fields}) and $\sigma_B$ is the error derived for the magnetic field strength in Ref. \cite{Zamaninasab_2014}\footnote{Errors are derived from their Eq. (4), and the errors on the values quoted in their Table 1 and references therein.}. In each case, $\sigma_B$ is around 20\% of $\Bar{B}_0$.
For each field realization, we fit the ALP spectrum to the data by varying $B_0$ and $\boldsymbol{\theta}$ in such a way as to maximize $\Lik_{\mathrm{ALP}}(m_a,g_{a\gamma},\mathbf{B}_j,\boldsymbol{\theta})$. We use the \texttt{iminuit} \texttt{PYTHON} package for the fitting. Note that every time $B_0$ is changed in the fit, $P_{\gamma\gamma}$ has to be recalculated completely (as is the case when changing $m_a$ or $g_{a\gamma}$). When $B_0$ changes, only the overall field strength is affected; the random field orientations and domain lengths remain the same for a given $j$. As shown in Appendix \ref{sec:appendix_seds}, changing the synchrotron photon field has a negligible effect on dispersion, so we can keep it constant as well. Large variations of $B_0$, such as removing the field completely, are discouraged by the prior term in the likelihood. Best-fit values of the field strength and spectral parameters are denoted by $\mathbf{\Hat{B}}_j$ and $\boldsymbol{\Hat{\theta}}$.\par
We scan an $8\times7$ logarithmically spaced grid in ($m_a,g_{a\gamma}$)-space, with $ m_a \in [5,5000]\text{ neV}$ and $g_{a\gamma} \in [0.1,10]\times 10^{-11}\text{ GeV}^{-1}$. This region is where we might expect mixing in the jets (see Ref. \cite{Davies_2021}), with the critical energies (Eqs. \eqref{eq:e_low} and \eqref{eq:e_high}) lying around the \textit{Fermi} energy range. For masses $m_a<5$ neV, the precise jet length becomes important, as does conversion in the IGMF. Higher couplings are ruled out by experiment \cite{CAST_2017}, and lower couplings would lead to oscillations too small to be detectable.\par
 In order to treat the random field statistically (following, e.g., Refs. \cite{Fermi_2016,CTA_Gpropa_2021}), for each ($m_a$, $g_{a\gamma}$) we perform the fits for 100 magnetic field realizations, then sort them by $\Lik_{\mathrm{ALP}}$ and choose $j=95$ corresponding to the 0.95 magnetic field realization quantile. Each point on the ALP grid then corresponds to a likelihood value, $\Lik^k_{\mathrm{ALP}}(m_a,g_{a\gamma},\mathbf{\Hat{B}}_{95},\boldsymbol{\Hat{\theta}})$, for each event type. \par
\begin{figure}
  \centering
    \includegraphics[width=0.48\textwidth]{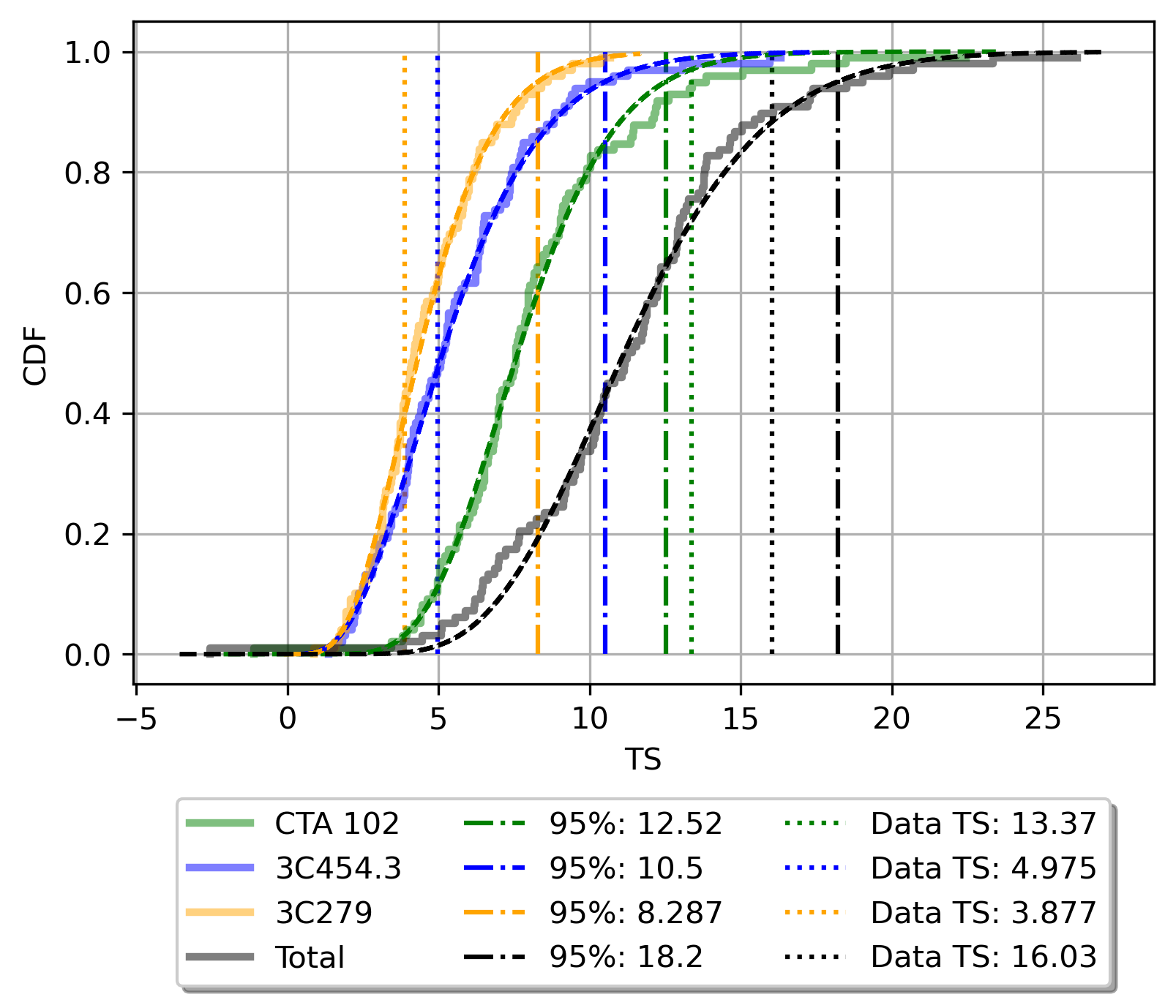}
  \caption{Cumulative distribution functions (CDF) for the $\mathrm{TS}$ values for the individual sources (solid), and the total $\mathrm{TS}_\mathrm{tot}$ distribution (black). Dashed lines show the best-fit gamma distributions. Dot-dashed vertical lines show the 95\% thresholds and dotted vertical lines show the $\mathrm{TS}$ values of the data.}
  \label{fig:indiv_null_dists}
\end{figure}

 For each source, the overall ALP and no-ALP hypotheses can be compared with the test statistic,
\begin{equation}\label{TS_indiv}
    \TS = -2\sum_k\ln\left(\frac{\Lik_0^k(\boldsymbol{\Bar{\theta}})}{\Lik_{\mathrm{ALP}}^k(\Hat{m}_a,\Hat{g}_{a\gamma},\mathbf{\Hat{\Hat{B}}}_{95},\boldsymbol{\Hat{\Hat{\theta}}})}\right),
\end{equation}
defined in the standard way for a likelihood ratio test, where $\Lik_0^k$ are the maximum likelihoods for the no-ALP model, found in Eq. \eqref{eq:L_0}, with best-fit spectral parameters, $\boldsymbol{\Bar{\theta}}$, and the additional hats in the denominator denote the maximum likelihood over the whole ALP grid \cite{Rolke_2005}. A high value of $\TS$ would mean that the ALP hypothesis is more likely than the no-ALP hypothesis. To quantify how confidently we could reject the no-ALP hypothesis for a given $\TS$, we use Monte Carlo simulations to find the null $\TS$ distribution. For this, we use the \texttt{simulate\_roi} function in \texttt{FERMIPY} to generate 100 simulated ROIs for each event type (for each of our sources), every time removing the source and injecting a new one from the best-fit spectral model, including photon absorption. The whole analysis is then repeated on each simulated ROI to get a distribution in $\TS$ (shown as the solid lines in Fig. \ref{fig:indiv_null_dists}). The $\TS$ thresholds can then be read from these distributions. Specifically, because we only have 100 simulations, we fit gamma distributions to the TS distributions (dashed lines), from which we can take the 0.95 threshold values (dot-dashed lines in Fig. \ref{fig:indiv_null_dists}). As can be seen, $\TS>10.5$ is required to reject the no-ALP hypothesis with 95\% confidence for 3C454.3; $\TS>12.52$ for CTA 102; and $\TS>8.29$ for 3C279.\par
\begin{figure}
  \centering
    \includegraphics[width=0.48\textwidth]{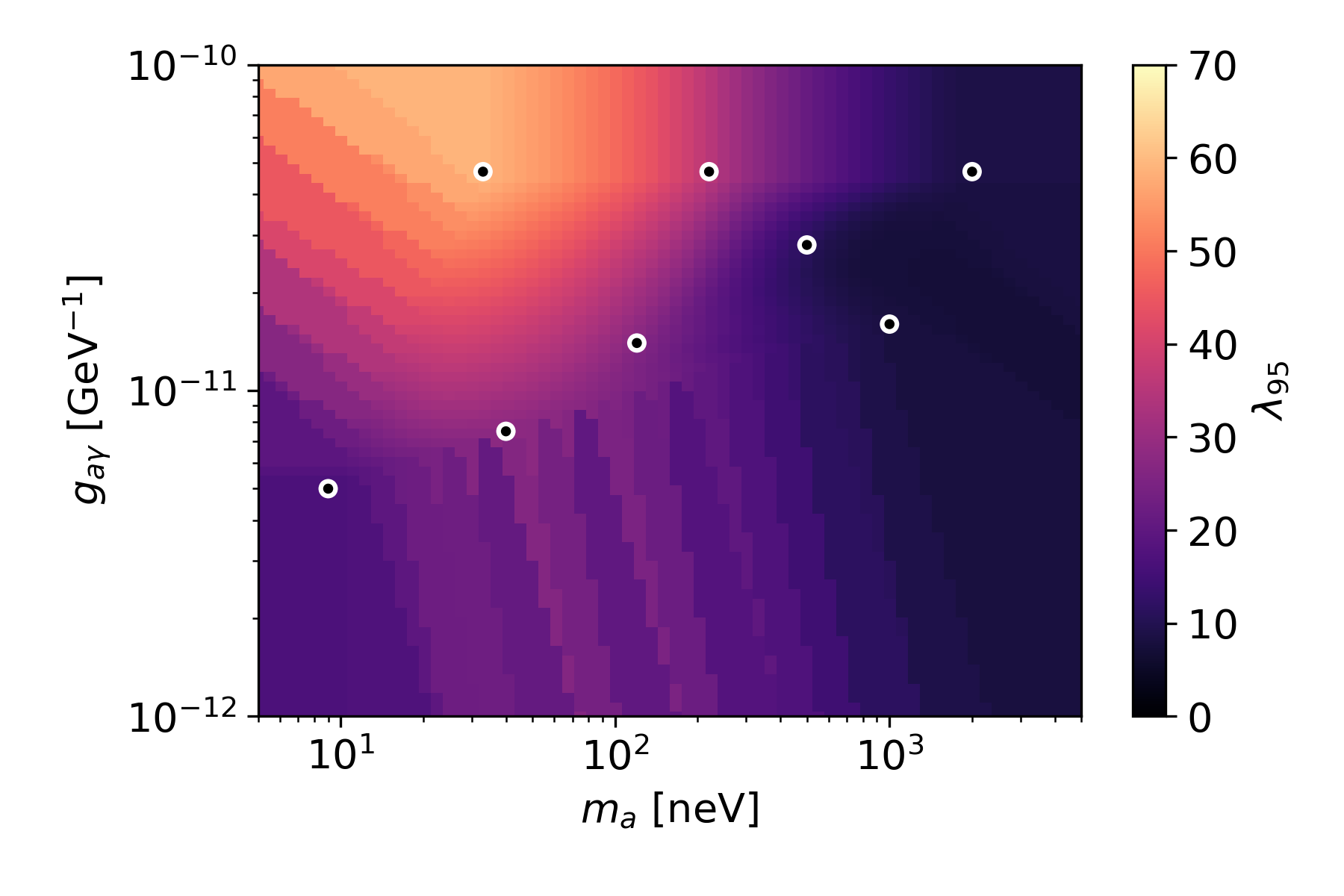}
    \includegraphics[width=0.48\textwidth]{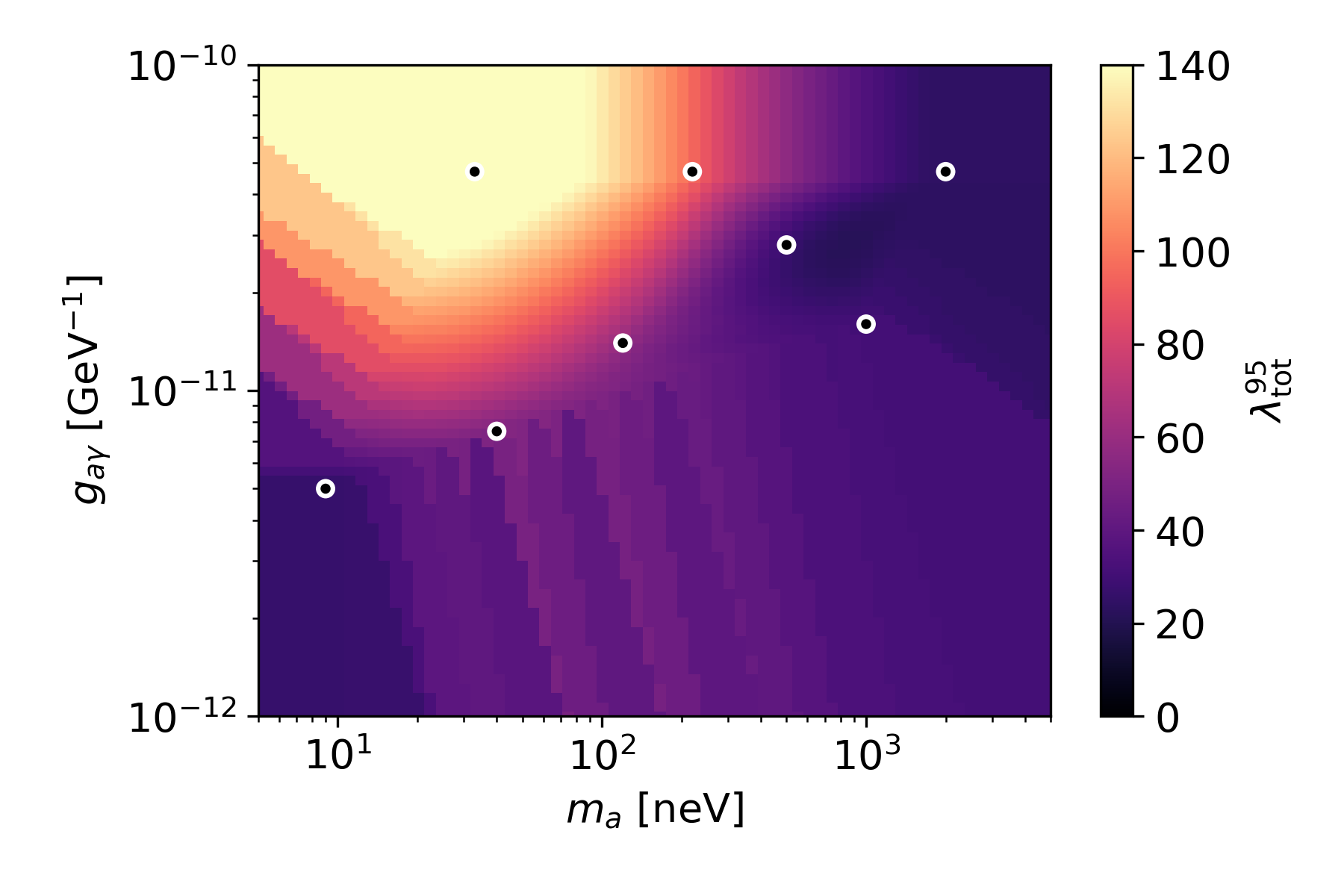}
  \caption{$\lambda_{95}$ thresholds for CTA 102 (top) and all sources ($\lambda_\mathrm{tot}$) (bottom). Points show injected ($m_a$,$g_{a\gamma}$) points, which are interpolated between. Outside the bounds of the injected points, the nearest neighbor is used for the threshold.}
  \label{fig:lambda_thresholds}
\end{figure}

Regardless of whether we can claim an ALP detection, we would also like to find which ALP parameters a given observation is inconsistent with---i.e., which values of $m_a$ and $g_{a\gamma}$ can be excluded. This can be done with a different test statistic:
\begin{equation}\label{Lambda_indiv}
    \lambda(m_a,g_{a\gamma}) = -2\sum_k\ln\left(\frac{\Lik_{\mathrm{ALP}}^k(m_a,g_{a\gamma},\mathbf{\Hat{B}}_{95},\boldsymbol{\Hat{\theta}})}{\Lik_{\mathrm{ALP}}^k(\Hat{m}_a,\Hat{g}_{a\gamma},\mathbf{\Hat{\Hat{B}}}_{95},\boldsymbol{\Hat{\Hat{\theta}}})}\right),
\end{equation}
which compares the best fit at each point ($m_a$, $g_{a\gamma}$) with the overall best fit of the whole grid. A large $\lambda$ means that the best fit at that point is significantly worse than the best fit overall and so can be rejected by the data.
The underlying distribution of $\lambda(m_a,g_{a\gamma})$ can be found in the same way as the $\TS$ distribution, except this time an ALP spectrum is injected into the simulated ROIs. This distribution could, in principle, be different for each point in ALP parameter space, as the oscillations do not depend trivially on $m_a$ and $g_{a\gamma}$. Doing the simulations for every point ($m_a$, $g_{a\gamma}$) is not computationally feasible, however.  Therefore we calculate the $\lambda$ distribution at various points and linearly interpolate between them to get the 95\% $\lambda$ thresholds across the grid, $\lambda_{95}(m_a,g_{a\gamma})$. For CTA 102, we use eight points, spread over the region of parameter space where we might expect exclusions; the top panel of Fig. \ref{fig:lambda_thresholds} shows $\lambda_{95}$ for CTA 102. The black points show the injected ($m_a$, $g_{a\gamma}$) pairs. Within the region bounded by these points, $\lambda_{95}$ is interpolated, and it can be seen that $\lambda_{95}$ is not constant, but varies smoothly across the grid---generally lowering for decreasing coupling and increasing mass, i.e., for weaker oscillations. Outside the region bounded by the injected points, nearest-neighbor $\lambda_{95}$ values are used. This is generally a conservative estimate, as $\lambda_{95}$ would continue to decrease into the unprobed parameter space beyond the lower right edge of the interpolation region, where it would approach the $\mathrm{TS}$ threshold. Because of the smoothness of the CTA 102 $\lambda_{95}$ distribution, we use fewer injected points for 3C454.3 and 3C279---seven and three respectively---to save on computing time.

So far, we have only discussed individual source analyzes. It is possible to combine the likelihoods from the different sources in the same way as those from the different event types within one source (or even different energy bins within one event type). The total ALP and no-ALP likelihoods are just the product of the individual source likelihoods, $\mathfrak{L}=\Pi_s \Lik_s$, where $s$ indexes the different sources. This means the final $\TS_{\mathrm{tot}}$ and $\lambda_{\mathrm{tot}}(m_a, g_{a\gamma})$ formulae are\footnote{Comparable to those used in, e.g., \textit{Fermi} searches for dark matter annihilation lines in dwarf spheroidal galaxies \cite{Fermi_DSph_2015} but with the $B$-field taking the role of the $J$-factor in our case.}
\begin{equation}\label{TS_tot}
    \TS_\mathrm{tot} = -2\sum_k\ln\left(\frac{\mathfrak{L}_0^k(\boldsymbol{\Bar{\theta}})}{\mathfrak{L}_{\mathrm{ALP}}^k(\Hat{m}_a,\Hat{g}_{a\gamma},\mathbf{\Hat{\Hat{B}}}_{95},\boldsymbol{\Hat{\Hat{\theta}}})}\right),
\end{equation}
and
\begin{equation}\label{eq:Lambda_tot}
    \lambda_\mathrm{tot}(m_a,g_{a\gamma}) = -2\sum_k\ln\left(\frac{\mathfrak{L}_{\mathrm{ALP}}^k(m_a,g_{a\gamma},\mathbf{\Hat{B}}_{95},\boldsymbol{\Hat{\theta}})}{\mathfrak{L}_{\mathrm{ALP}}^k(\Hat{m}_a,\Hat{g}_{a\gamma},\mathbf{\Hat{\Hat{B}}}_{95},\boldsymbol{\Hat{\Hat{\theta}}})}\right),
\end{equation}
where the minima are found after the product over the sources is taken. Of course, because the intrinsic parameters of each source are different, the different sources will be capable of probing slightly different regions of parameter space to greater or lesser degrees. It is important that the likelihoods are combined in this way so that each source contributes proportionally to the overall likelihood.
The distributions for these two test statistics can be found in the same way as those for the individual sources. Figure \ref{fig:indiv_null_dists} also shows the $\TS_\mathrm{tot}$ distribution; a value $\TS_\mathrm{tot}>18.2$ would be required to reject the no-ALP hypothesis with 95\% confidence for all the sources combined. The lower panel of Fig. \ref{fig:lambda_thresholds} shows the $\lambda_\mathrm{tot}^{95}$ thresholds across the ALP parameter space. The same interpolation method is used as before, and, again, the overall distribution varies smoothly. For those points where either one or both of 3C454.3 and 3C279 is missing injected simulations, we use the sum of the individual $\lambda_{95}$ thresholds. This again is a conservative (i.e., over-) estimate of $\lambda_\mathrm{tot}^{95}$, as by definition $\lambda_\mathrm{tot}^{95} \leq \sum_s \lambda_{95}^s$ everywhere (the two are only equal if the minima of the likelihood profiles for all the sources lie at the same point). Only the point at 1000 neV does not have injected simulations for both CTA 102 and 3C454.3, which together should dominate the overall thresholds, so in the relevant region of parameter space this approximation is small.
\begin{figure}
  \centering
    \includegraphics[width=0.48\textwidth]{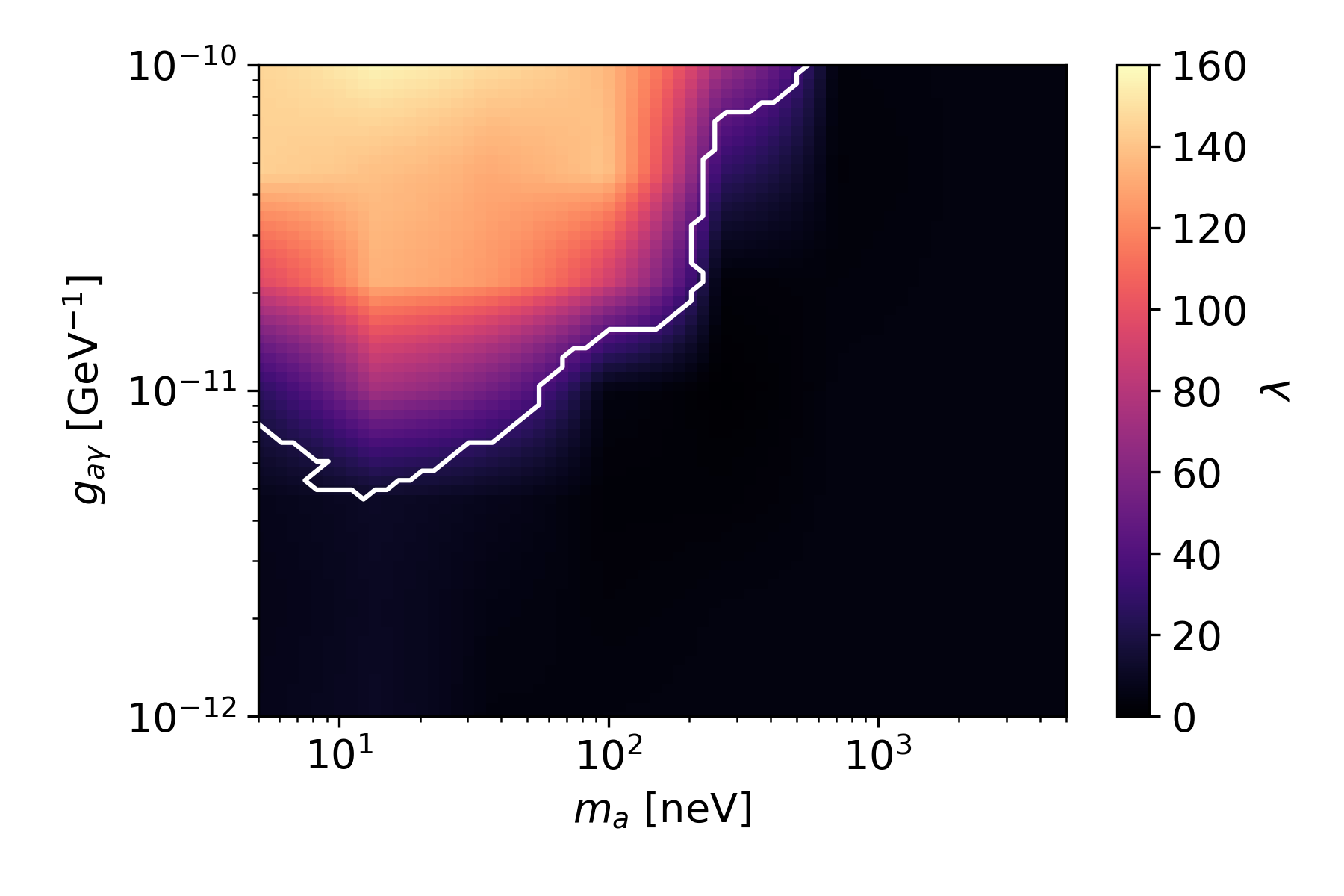}
    \includegraphics[width=0.48\textwidth]{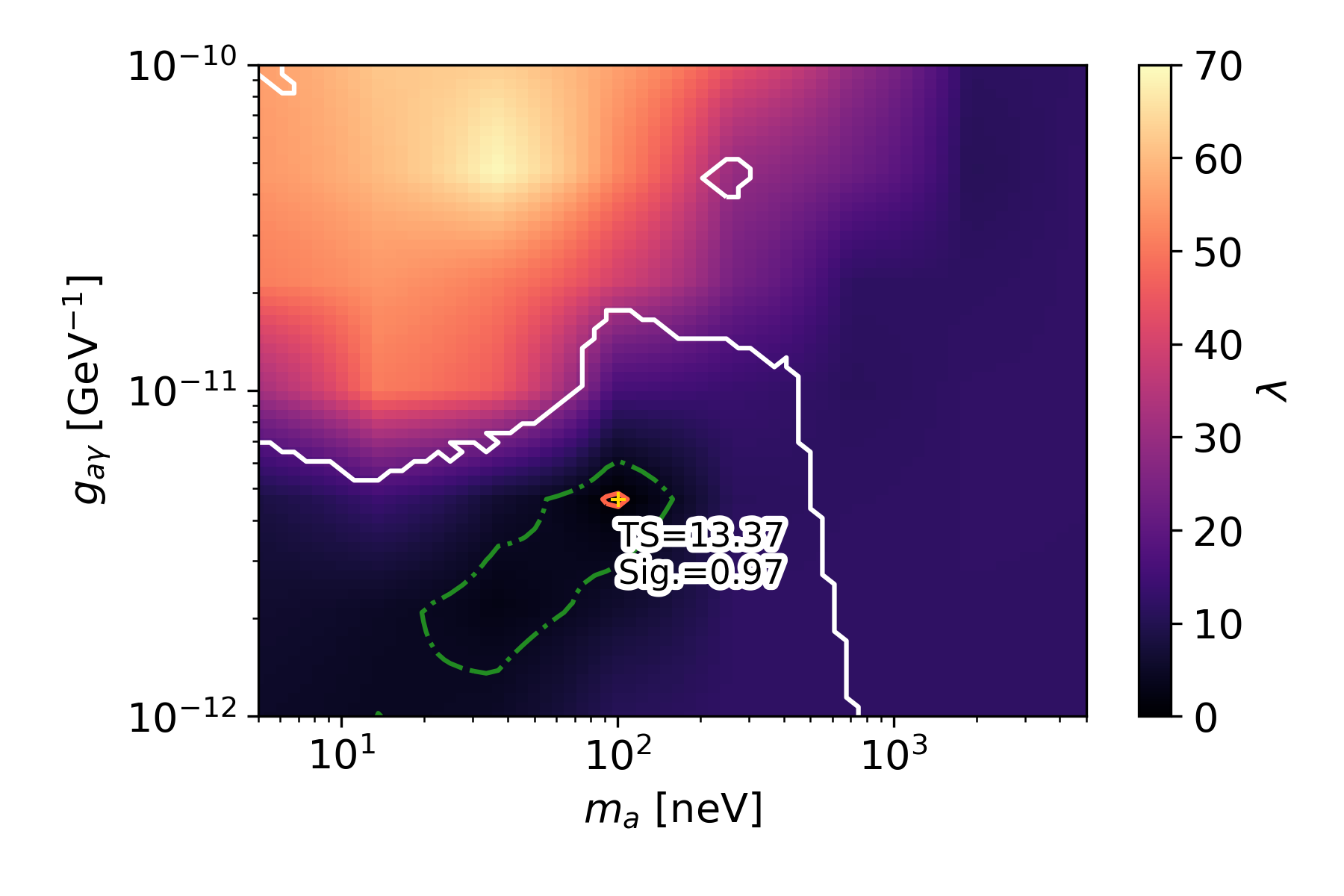}
    \includegraphics[width=0.48\textwidth]{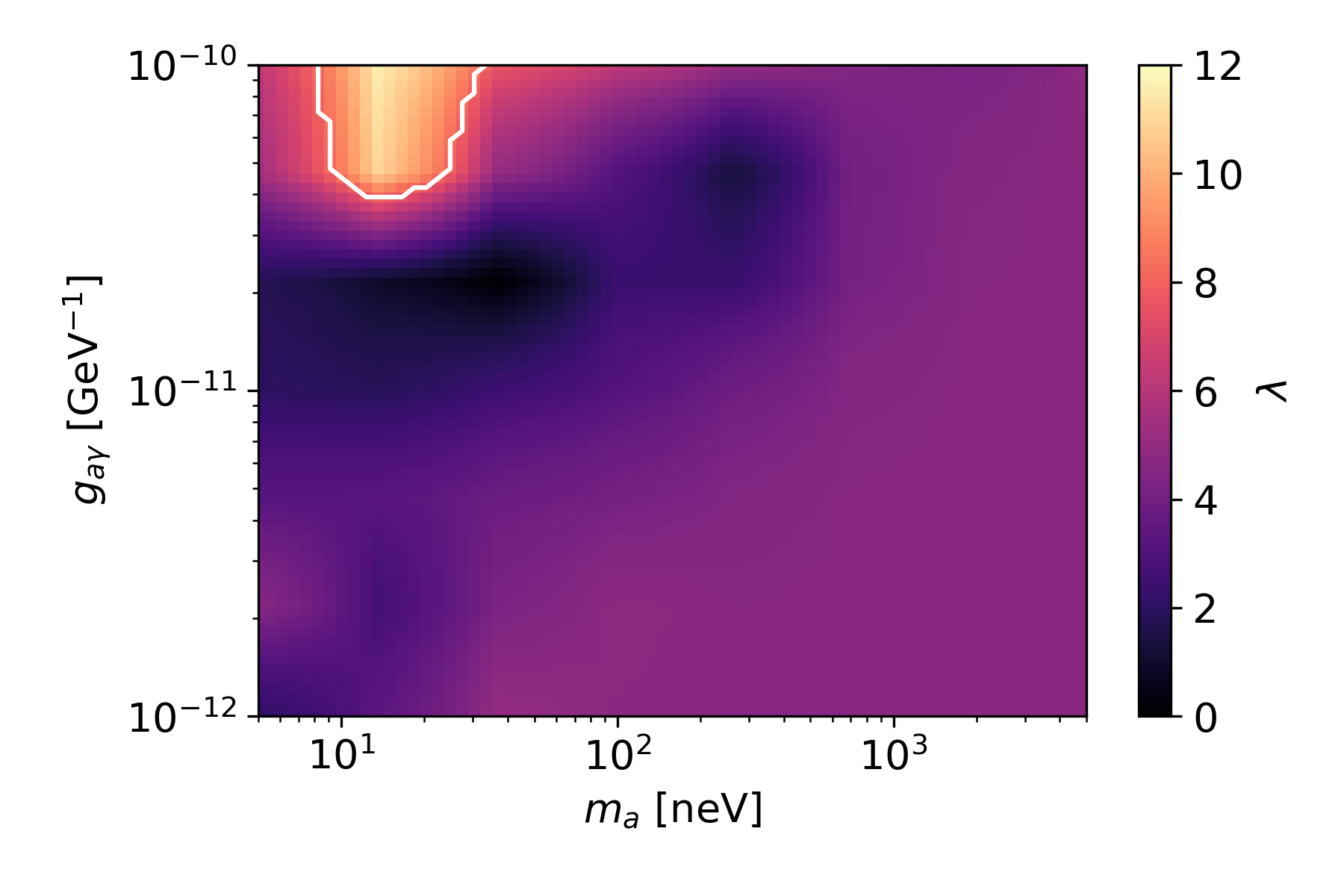}
  \caption{$\lambda(m_a,g_{a\gamma})$ for 3C454.3 (top), CTA 102 (middle), and 3C279 (bottom). White contours show the 95\% exclusions, i.e., they enclose regions where $\lambda\geq\lambda_{95}$. The green dot-dashed and the red solid contours show the $1\sigma$ and $2\sigma$ preference regions for CTA 102 respectively, and the gold cross shows the location of the best-fit point.}
  \label{fig:indiv_constraints}
\end{figure}
\section{Results} \label{sec:results}
Figure \ref{fig:indiv_null_dists} also shows (dotted vertical lines) the data $\TS$ values for all the sources both individually and in combination. The $\TS$ values for 3C454.3 and 3C279 are below their respective $\TS$ thresholds: $\mathrm{TS}_\mathrm{3C454.3}=4.97$, and $\mathrm{TS}_\mathrm{3C279}=3.88$. For CTA 102, however, their is a slight preference ($2\sigma$) for the ALP case\footnote{For comparison, a $5\sigma$ significance is generally required for a new particle detection within the particle physics community.}: $\mathrm{TS}_\mathrm{CTA 102}=13.37$, which is over the threshold of $12.52$. $\mathrm{TS}=13.37$ is in the 97\% quantile of the gamma-function fitted to the CTA 102 $\mathrm{TS}$ distribution, but falls to the 91\% quantile if the distribution is simply read from the simulations. Also, this local significance of $\sim 2\sigma$ for an ALP signal in the CTA 102 data would be further reduced by a trial factor of 3 when considering the fact we looked at three sources. Therefore, this is not a very significant preference for the ALP case, and indeed, it disappears in the combined analysis: $\mathrm{TS}_\mathrm{tot}=16.03$. Overall then, we cannot rule out the no-ALP hypothesis, or in other words, we have not found an ALP signal in the data.\par
Nonetheless, we are able to place limits on the parameters $m_a$ and $g_{a\gamma}$. Figure \ref{fig:indiv_constraints} shows $\lambda$ for each of the sources. The white contours enclose regions where $\lambda\geq\lambda_{95}$, and so show the 95\% exclusion contours for each individual source. For CTA 102, the regions of $1\sigma$ (68\%) and $2\sigma$ (95\%) preference over the null hypothesis are also shown as green dot-dashed and red solid contours respectively. The best-fit point ($m_a=100.8$ neV and $g_{a\gamma}=4.64\times10^{-12}$ GeV$^{-1}$) is also plotted as a gold cross, along with its significance ($0.97$; $2.17\sigma$). As can be seen, because of this slight preference for the ALP case, the 95\% exclusions contour from the CTA 102 data extend to high masses and low couplings (which approximates the no-ALP case).

Aside from CTA102, the constraints from 3C454.3 are the strongest, as would be expected from the comparatively good statistics of the 3C454.3 observations (see Fig. \ref{fig:seds}). Constraints from 3C279 data are much weaker than the other sources not only because its statistics are not quite as good, but also because the configuration of the field parameters means that, with $B_0$ free, good fits are generally able to be found to the data; for 3C279, $\lambda$ is smaller for much of the region that is excluded by the other sources than it is in the high-mass--low-coupling region. This highlights the importance of leaving the magnetic field strength free in the fitting. \par

\begin{figure}
  \centering
    \includegraphics[width=0.48\textwidth]{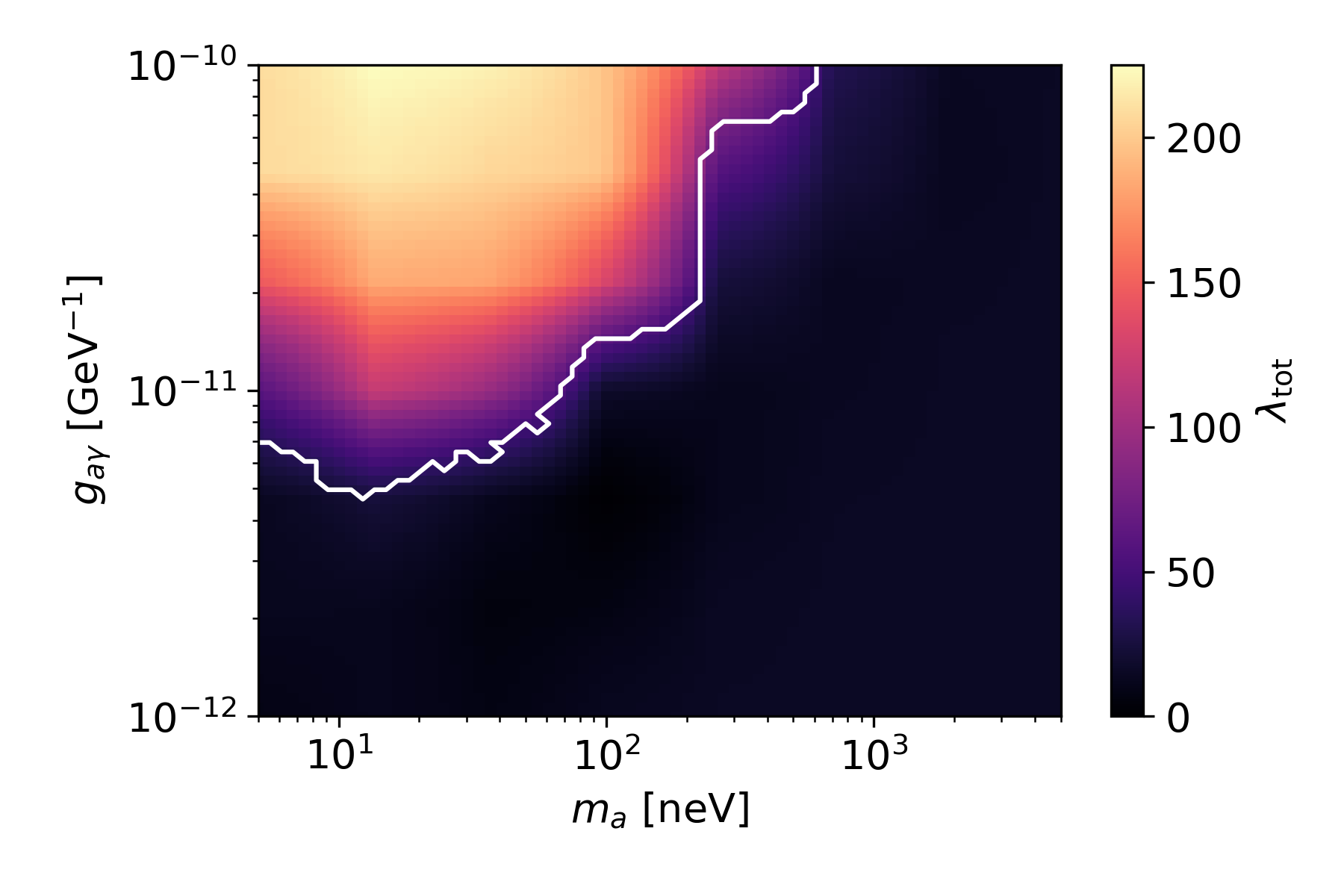}
  \caption{$\lambda_\mathrm{tot}(m_a,g_{a\gamma})$ for all the sources combined. White contour shows the 95\% exclusions, i.e., enclosing the region where $\lambda_\mathrm{tot}\geq\lambda_\mathrm{tot}^{95}$.}
  \label{fig:combined_constraints}
\end{figure}
\begin{figure}
  \centering
    \includegraphics[width=0.48\textwidth]{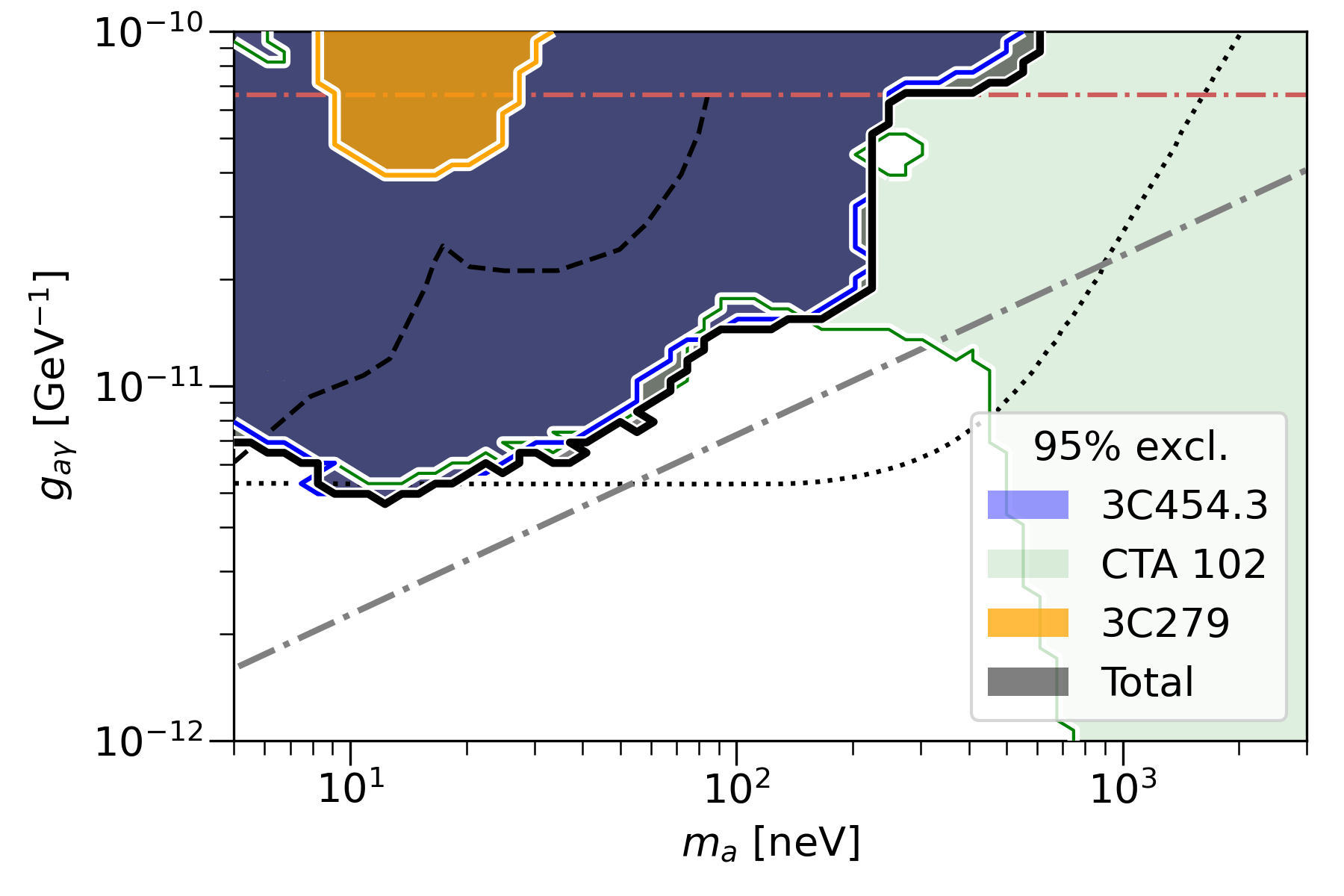}
  \caption{Overall 95\% exclusion contours for each source and for the combined analysis. The black dotted contour shows constraints from magnetic white dwarf radio polarization \cite{Dessert_Dunsky_Safdi_2022}. Black dashed contours show previous gamma-ray constraints \cite{HESS_2013,Fermi_2016}. The red dot-dash contour shows the CERN Axion Solar Telescope (CAST) experimental constraints \cite{CAST_2017}. The dark matter line is shown in grey dot-dash.}
  \label{fig:filled_constraints}
\end{figure}

As was shown in Fig. \ref{fig:indiv_null_dists}, the preference for the ALP case shown in the CTA 102 data disappears in the combined analysis; we would therefore expect 3C454.3 to contribute most strongly to the combined exclusions. Indeed, Fig. \ref{fig:combined_constraints} shows $\lambda_\mathrm{tot}$ for the whole scanned parameter space, and the 95\% exclusions (again shown by the white contour) are only marginally better than the 3C454.3 exclusions. In particular, the high-mass--low-coupling region is not excluded. This highlights the benefits of using a combined analysis to derive robust exclusions.
Figure \ref{fig:filled_constraints} shows how our 95\% exclusion contours compare with current constraints, shown by the black and red dashed contours. The dark matter line is shown as a grey dot-dashed line, below which ALPs could make up all of dark matter \cite{Arias_2012}. As can be seen from the figure, the combined analysis performed here allows the previous gamma-ray constraints to be extended.
Overall, we can exclude the parameter space $5\mathrm{neV}\lesssim m_a\lesssim200$ neV and $g_{a\gamma}\gtrsim 5 \times 10^{-12}$ GeV$^{-1}$ with 95\% confidence.
\section{Conclusions} \label{sec:conclusions}
Searches for ALP signals in the high energy spectra of AGN have provided some of the strongest constraints on ALP $m_a$ and $g_{a\gamma}$ so far \cite{HESS_2013,Fermi_2016,Reynolds_2020}.
These searches have all been analyzes of individual sources and have generally used the turbulent magnetic field of the host cluster as their main mixing region; similar searches are also planned in the future (e.g, \cite{CTA_Gpropa_2021}).\par
Here, we have performed, for the first time, a combined analysis on \textit{Fermi}-LAT data of three bright, flaring FSRQs (3C454.3, CTA 102, and 3C279), with the blazar jets themselves as the dominant mixing region. These sources were chosen because they displayed the brightest flaring periods over the \textit{Fermi} lifetime.
\par
We analyze each of the sources using the \texttt{FERMIPY} \texttt{PYTHON} package, first over a significant fraction of the \textit{Fermi} lifetime to get average ROI models which are then used as initial conditions for detailed SED analysis of the flaring time periods. This enables us to extract likelihood curves from the resulting flare SEDs, which can be used to compare ALP spectral models to the data with a log-likelihood ratio test.\par
To find the ALP spectra, we model the jets within the PC framework, with a helical and a tangled field component as outlined in Ref. \cite{Davies_2021}. In particular, based on observed polarization fractions of the sources, we use jets with 30\% of the magnetic energy density in the tangled component. Also, for the first time, we include a full treatment of photon-photon dispersion within the jet, following Ref. \cite{Davies_2022}. This requires the modeling of the disk, BLR, torus, synchrotron, starlight, CMB and EBL photon fields within the jets. We have performed SED modeling with our combined jet and photon-field models to ensure they are consistent with both each other and observations.\par
These jet models then allow us to compute ALP spectra for each of the sources and fit them to the \textit{Fermi} observations. We treat both the tangled field component and the errors in the data statistically, by running the analysis for 100 field simulations on 100 simulated \textit{Fermi} data sets. Also, unlike previous work, we account for the uncertainty in our $B$-field model by leaving the field strength free in the fits, including a prior term in the likelihood function based on core-shift estimates of the field strength.\par
To find the underlying distributions of the test statistics used to place limits, we performed the analysis on simulated data with various ALP-spectra injected into it. This was done across the ALP parameter space, enabling a 2D test-statistic threshold to be constructed, as opposed to using a single value everywhere.\par
In the CTA 102 data, we find a marginal ($2\sigma$) preference for ALPs, with the best fit occurring at $m_a=100.8$ neV and $g_{a\gamma}=4.64\times10^{-12}$ GeV$^{-1}$ (below the dark matter line). This slight preference disappears in the combined analysis, however, highlighting the benefits of using multiple sources.
Overall then, we find no evidence for ALPs, but are able to exclude the parameter space $5 \mathrm{ neV}\lesssim m_a\lesssim200$ neV and $g_{a\gamma}\gtrsim 5 \times 10^{-12}$ GeV$^{-1}$ with 95\% confidence. This is an improvement on previous gamma-ray searches in this mass range, though it is almost completely contained within the magnetic white dwarf polarization constraints of \cite{Dessert_Dunsky_Safdi_2022}. Our constraints do not quite reach the dark matter line, but are limited in coupling to similar $g_{a\gamma}$ values as previous \textit{Fermi} limits (see \cite{Fermi_2016}), which is to be expected. Nonetheless, we reach lower couplings than those projected to be reached by the future ALPS II experiment in the same mass range \cite{Ortiz_ALPsII_2020}, and comparable couplings to the projected limits of the future IAXO experiment \cite{IAXO_2014}. \par
Future searches, with CTA for instance (like those outlined in \cite{CTA_Gpropa_2021}), could likely take advantage of greater instrumental sensitivity to probe lower couplings using this same method, with the blazar jets as the dominant mixing region. It would also be interesting to see how these limits could be extended in the event of another flare as bright as the one from 3C454.3 used here, to fully take advantage of the combined analysis method.
\section*{Acknowledgements} \label{sec:acknowledgements}
 M.~M.  acknowledges  support from the European Research Council (ERC) under the European Union's Horizon 2020 research and innovation program Grant agreement No. 948689 (AxionDM) and from the Deutsche Forschungsgemeinschaft (DFG, German Research Foundation) under Germany's Excellence Strategy – EXC 2121 „Quantum Universe" – 390833306.
The \textit{Fermi} LAT Collaboration acknowledges generous ongoing support
from a number of agencies and institutes that have supported both the
development and the operation of the LAT as well as scientific data analysis.
These include the National Aeronautics and Space Administration and the
Department of Energy in the United States, the Commissariat \`a l'Energie Atomique
and the center National de la Recherche Scientifique / Institut National de Physique
Nucl\'eaire et de Physique des Particules in France, the Agenzia Spaziale Italiana
and the Istituto Nazionale di Fisica Nucleare in Italy, the Ministry of Education,
Culture, Sports, Science and Technology (MEXT), High Energy Accelerator Research
Organization (KEK) and Japan Aerospace Exploration Agency (JAXA) in Japan, and
the K.~A.~Wallenberg Foundation, the Swedish Research Council and the
Swedish National Space Board in Sweden.

Additional support for science analysis during the operations phase is gratefully
acknowledged from the Istituto Nazionale di Astrofisica in Italy and the center
National d'\'Etudes Spatiales in France. This work performed in part under DOE
Contract DE-AC02-76SF00515.

\bibliographystyle{unsrt}

\appendix
\section{Field structure parameters}
\label{sec:appendix_params}
Ideally, all the field parameters ($f$, $\alpha$, $r_T$) would be left free in the fit, in the same way as $B_0$. Unfortunately, this is not computationally feasible at the moment ($P_{\gamma\gamma}$ has to be recalculated every time a field parameter changes in the fit). We therefore would like to constrain $f$, $\alpha$, and $r_T$ to specific values. \par
For our sources, at \textit{Fermi} energies, the fraction of magnetic energy density in the tangled component, $f$, has the strongest effect on the oscillations. In particular, a very low value of $f <0.05$ can greatly reduce the magnitude and severity of the oscillations produced by mixing in the jet. Fortunately, $f$ can be somewhat constrained on a source-by-source basis from radio polarization observations. In general, the fractional polarization of radio emission from a source is higher for a uniform field, and lower for a disordered field. Reference \cite{Zamaninasab_2013} compare Very-Long-Baseline Array fractional polarization maps of 3C454.3 to simulations, using a uniform helical field. They find that the asymmetry of the maps matches a helical field well, but an additional disordered field component is required to reduce the overall fractional polarization to observed levels. In particular, they need $B_t^2\sim0.45 B_h^2$, which equates to $f\sim0.3$. Fractional polarization maps of CTA 102 (see \cite{Li_2018}) look similarly asymmetric, and are at similar values. Reference \cite{Homan_2009} has done similar modeling, but with circular polarization as well, for 3C279. All of their best-fit models have $0.19 \leq f \leq 0.51$, again ruling out very low tangled field fractions, and their best fit value is $f=0.36$. Motivated by these results, we choose a fixed value, $f=0.3$, for all our sources.\par
\begin{figure}
  \centering
    \includegraphics[width=0.48\textwidth]{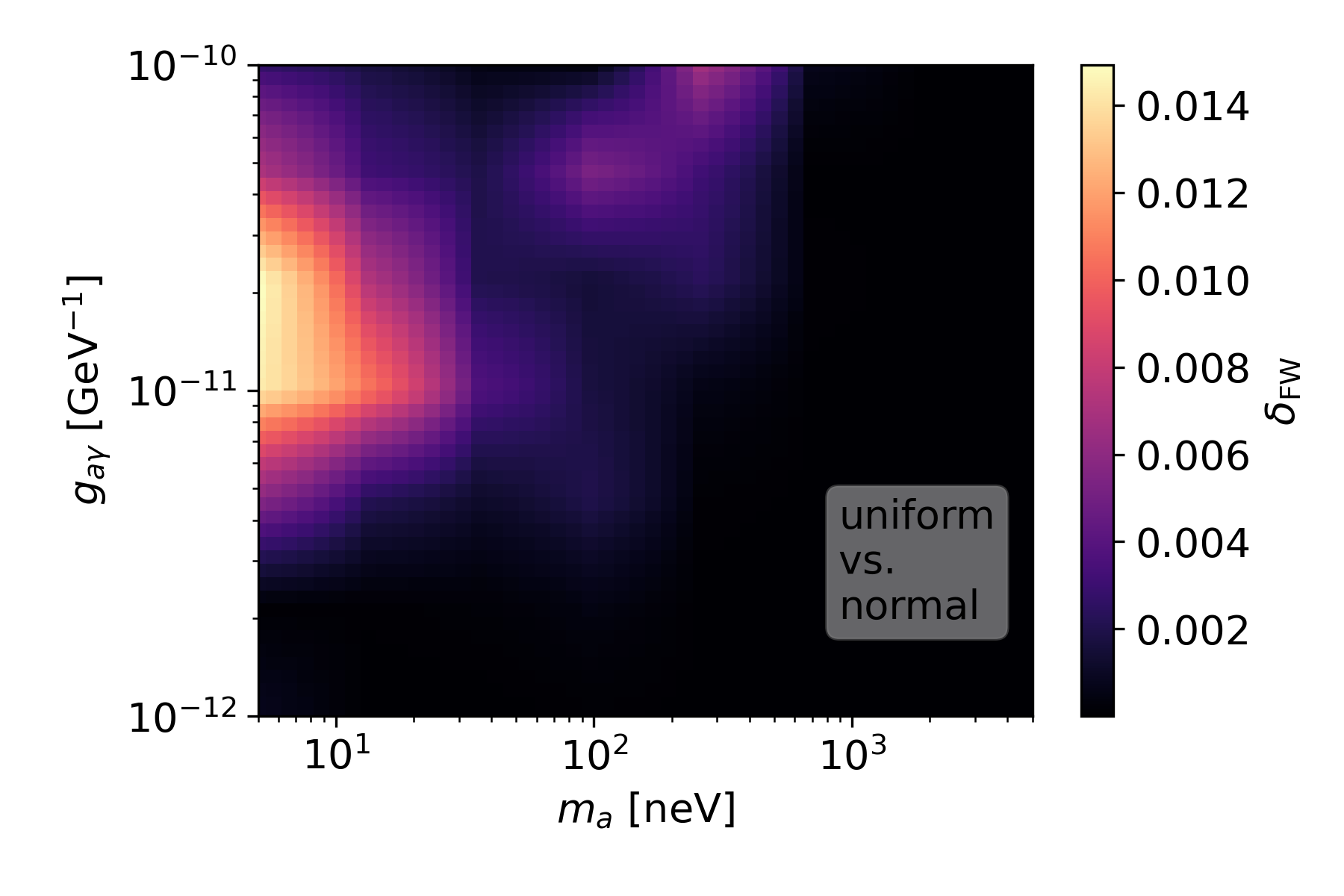}
  \caption{Average best-fit $P_{\gamma\gamma}$ differences (weighted by \textit{Fermi} counts), $\delta_\mathrm{FW}$, for 3C454.3, over 50 realizations between a uniform distribution of tangled coherence lengths ($l_c$) and a normal distribution. Differences are $< 2.5\%$.}
  \label{fig:jwdist_diffs}
\end{figure}

\begin{figure}
  \centering
    \includegraphics[width=0.48\textwidth]{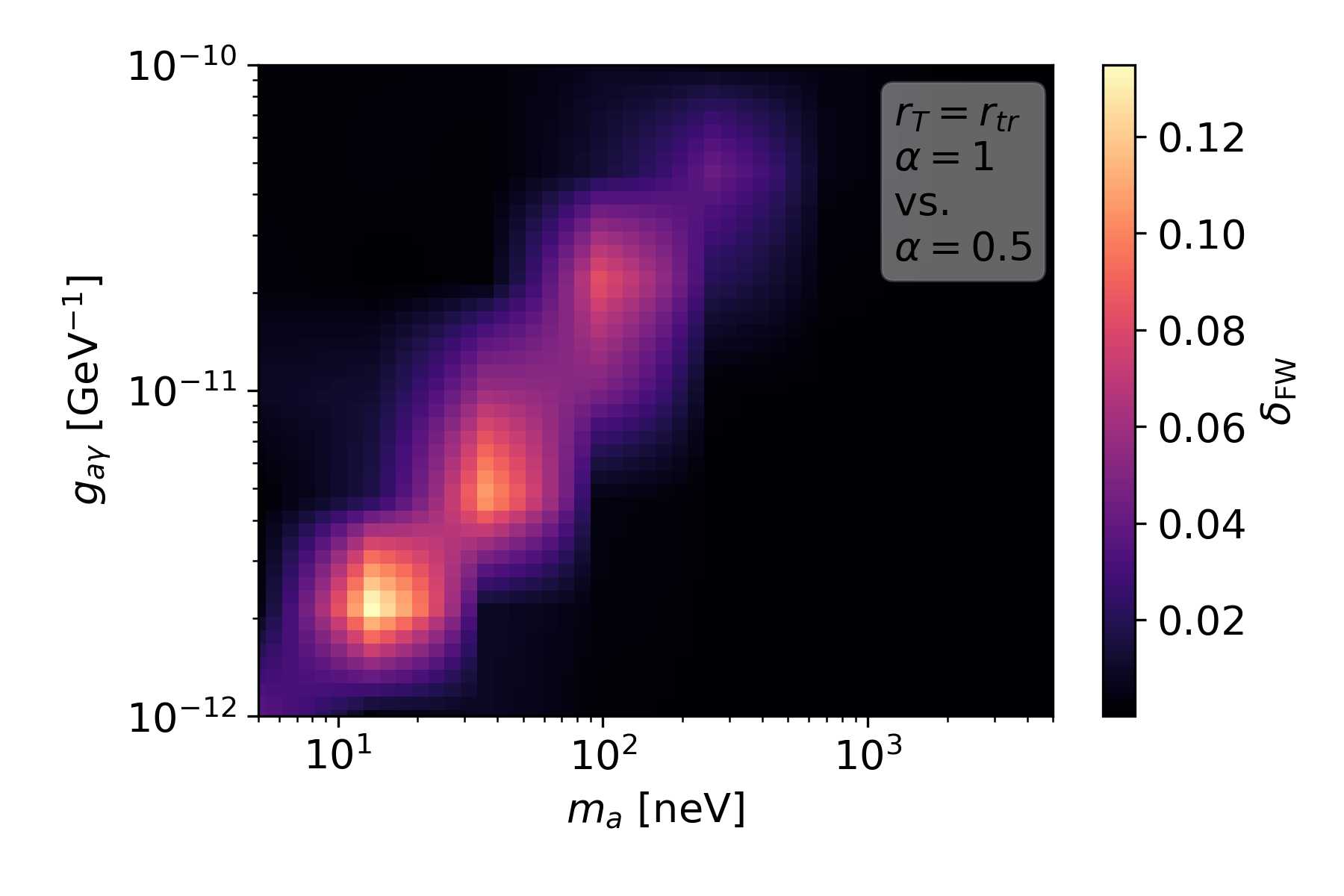}
  \caption{Average best-fit $P_{\gamma\gamma}$ differences (weighted by \textit{Fermi} counts), $\delta_\mathrm{FW}$, for 3C454.3, over 50 realizations for different jet field parameters: $\alpha=0.5$ and $\alpha=1$ with $r_T=r_\mathrm{tr}$ (top).}
  \label{fig:rT_alpha_diffs}
\end{figure}
Another factor concerning the tangled field component is which coherence length, $l_{c}$ to use. We take the jet width, $R$ as an upper limit for the tangled coherence length at a given $r$. The length of each tangled domain can then be drawn from a distribution of lengths less than the jet width. To investigate the effects of changing this distribution, we calculate best-fit $P_{\gamma\gamma}$s for 50 realizations of the tangled field (fixing $\alpha=1$ and $r_T=r_\mathrm{tr}$) in the jet of 3C454.3, for four different $l_c$ distributions: uniform, normal (with $\langle l_c \rangle=R/2$), linearly ascending ($\propto l_c$), and linearly descending ($\propto l_c^{-1}$). Over the $N_\mathrm{r}=50$ realizations, the average $P_{\gamma\gamma}$ differences between a uniform $l_c$ distribution and the others, weighted by \textit{Fermi} expected counts,
\begin{figure}
  \centering
    \includegraphics[width=0.48\textwidth]{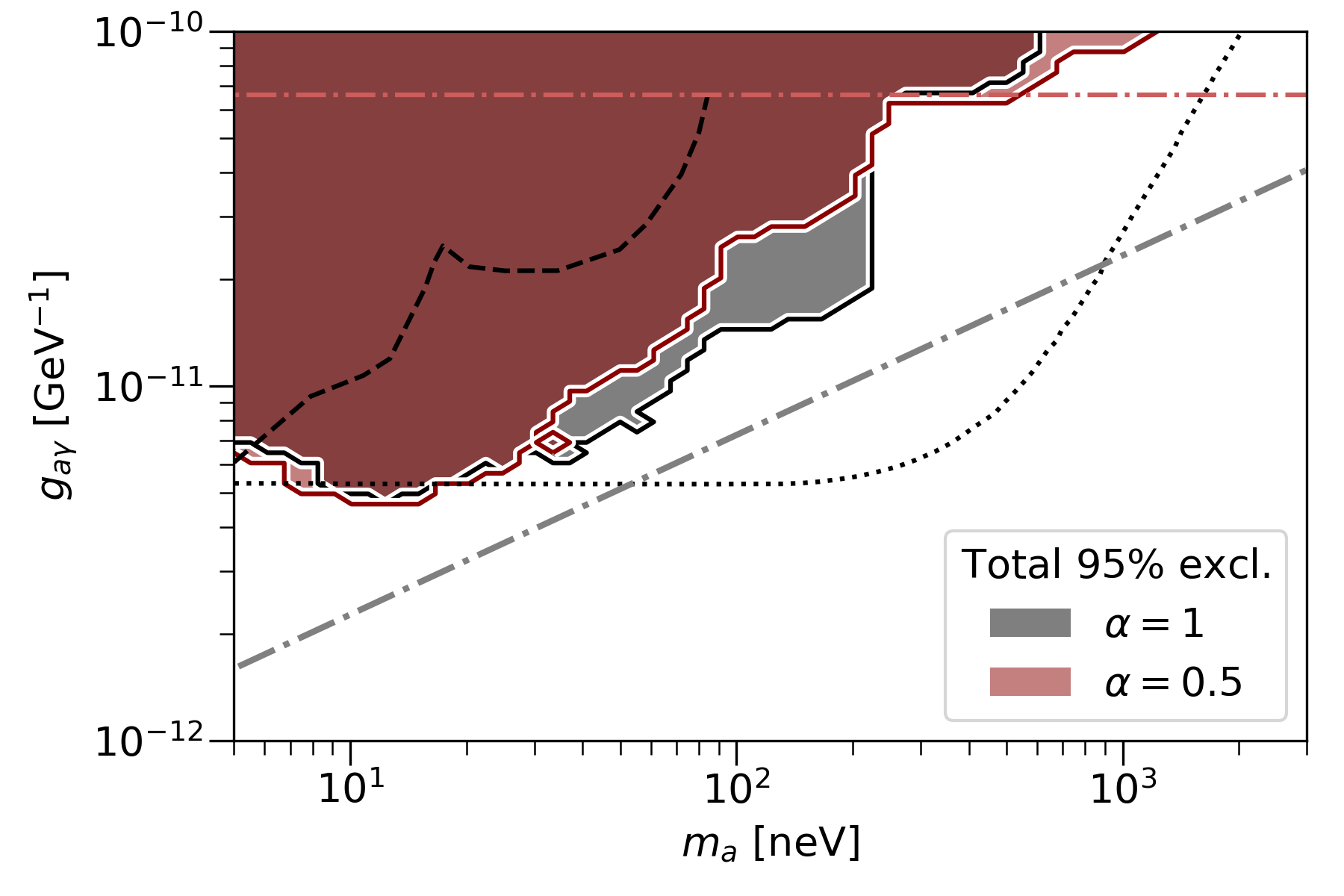}
  \caption{Overall 95\% exclusion contours for the combined analysis (black), and the same when $\alpha=0.5$ (red). The black dotted contour shows constraints from magnetic white dwarf radio polarization \cite{Dessert_Dunsky_Safdi_2022}. Black dashed contours show previous gamma-ray constraints \cite{HESS_2013,Fermi_2016}. The red dot-dash contour shows the CAST experimental constraints \cite{CAST_2017}. The dark matter line is shown in grey dot-dash.}
  \label{fig:filled_constraints_alpha}
\end{figure}
\begin{equation}\label{eq:FW_pgg_diffs}
    \delta_\mathrm{FW} = \frac{1}{N_\mathrm{r}}\sum_j\frac{\sum_i |\Delta P_{\gamma\gamma}(i,j)| \mu_i}{\sum_i \mu_i},
\end{equation}
 is below $2.5\%$ for all sources. Figure \ref{fig:jwdist_diffs} shows these differences between the uniform and the normal distributions; differences between all the other distributions look similar. This means that the differences between separate realizations of the tangled field outweigh the differences between specific distributions of tangled coherence lengths. We therefore do not need to model the $l_c$ distribution in detail, and can fix it to a uniform distribution with $l_c<R$.
\par The precise values of $r_T$ and $\alpha$ are hard to constrain observationally, but we can test their effects on the oscillations with the same average $P_{\gamma\gamma}$ method. When $\alpha=1$ is fixed, the differences between $r_T = 0.3$ pc and $r_T=r_\mathrm{tr}$ (again for 3C454.3) are below $1.5\%$. It seems that the differences in the ordered component of the field produced by varying $r_T$ are swamped by the differences between separate realizations of the tangled component, and also, $B_0$ can vary in the fit to compensate for any changes. Therefore it is reasonable to take $r_T=r_\mathrm{tr}$ for each of the three sources. Figure \ref{fig:rT_alpha_diffs} shows $\delta_\mathrm{FW}$ for $\alpha=0.5$ and $\alpha=1$ with $r_T=r_\mathrm{tr}$, when $B_0$ is left free in the fit, again for 3C454.3. The differences in this case can be larger ($\sim 10 \%$), but only in a few isolated regions. Note that $\alpha$ has a larger effect than $r_T$ because varying it can produce a larger change in the transverse field strength at the emission region. It is, of course, possible that these relatively large percentage differences in the $P_{\gamma\gamma}$s will not greatly affect the final results because they do not occur in important regions of parameter space. We perform the analysis using $\alpha=1$ throughout. In order to test the effects of changing $\alpha$ on our final results, we perform a single analysis of 3C454.3 and CTA 102 with $\alpha=0.5$, calculating $\lambda(m_a,g_{a\gamma})$ in the same way as described in Sec. \ref{sec:stat}. Figure \ref{fig:filled_constraints_alpha} shows how the total (combined) exclusions would vary in this case, using the same $\lambda_\mathrm{tot}^{95}$ threshold calculated with the ordinary analysis. As can be seen, the overall 95\% exclusions are only slightly changed by changing $\alpha$ to 0.5; in some places the exclusions are slightly better, in some places they are slightly worse. This is because the regions where $\alpha$ can make a large difference to the $P_{\gamma\gamma}$s are generally beyond, or at the edge of, our exclusion region. This means that, particularly at lower masses ($m_a\lesssim 200$ neV), our new exclusions are robust despite the approximations made concerning the magnetic field structure.\par
Overall, then, the values we choose are $\alpha=1$ and $r_T = r_\mathrm{tr}$ for all our sources.
Figure \ref{fig:Bs_example} shows one example field realization for 3C454.3 with $f=0.3$, $\alpha=1$ and $r_T=r_{tr}=59.8$ pc.
\section{Self-consistency of field and jet models}
\label{sec:appendix_seds}

 \begin{figure}[t]
  \centering
    \includegraphics[width=0.48\textwidth]{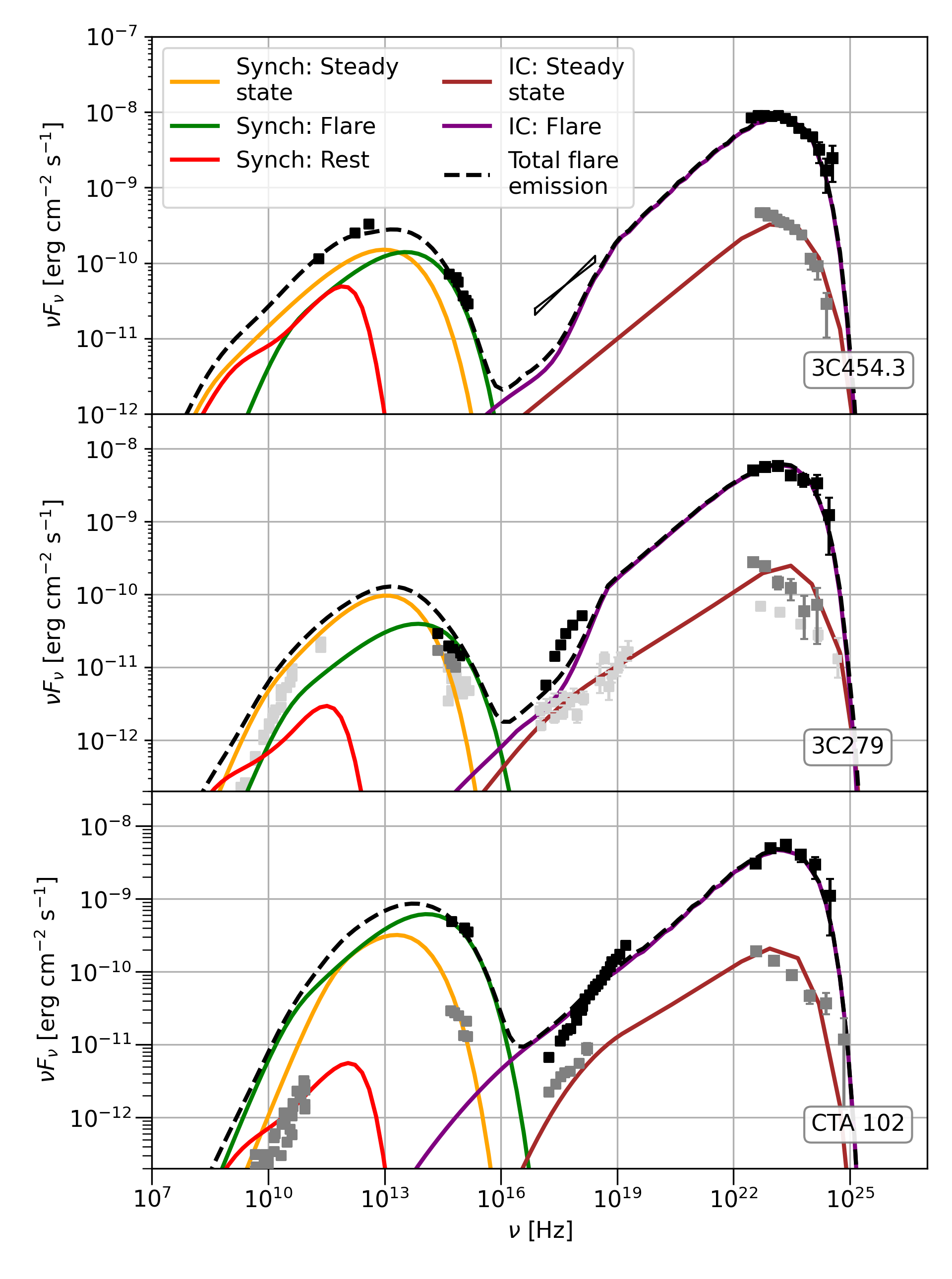}
  \caption{Modeled SEDs for our sources during flare and steady states: 3C454.3 (top), 3C279 (middle), CTA 102 (bottom). Data for 3C454.3 taken from \cite{Cerruti_2013}, for 3C279 from \cite{Potter_Cotter_NC_2015,HESS_3C279_1415_2019}, and for CTA 102 from \cite{Gasparyan_CTA102_2018}.}
  \label{fig:model_seds}
\end{figure}

  \begin{table}
 \caption{Parameters used for the blobs down the jet. The steady-state and flaring gamma-ray emission is produced from blobs at $r_{\mathrm{vhe}}$ and $r_{\mathrm{em}}$ respectively.}
  \centering
  \begin{tabular}{m{0.1\textwidth} m{0.12\textwidth} m{0.12\textwidth}m{0.1\textwidth}}
    \toprule\toprule
    Parameter & $r_{\mathrm{ss}}$ & $r_{\mathrm{em}}$  &
    Rest of jet\\
    \midrule
    \multicolumn{4}{c}{3C454.3} \\
    $r$ (pc) & 59.8 & 0.103 &   \\
    $B$ (G) & 0.013 & 1.9 & \\
    $n_e$ (cm$^{-3}$) & 4.71  & 7.9$\times$10$^3$ &  \\
    $E_c$ (MeV) & 1.3$\times10^3$ & 250 & 1.3 \\
    $\beta$ & 1.95 & 2 & 2  \\
    $\eta_\mathrm{em}$ &  & 2.8 &  \\
    \midrule
    \multicolumn{4}{c}{CTA 102} \\
    $r$ (pc) & 56.6 & 0.104 &   \\
    $B$ (G) & 0.026 & 3.64 & \\
    $n_e$ (cm$^{-3}$) & 2.5  & 3.7$\times$10$^3$ &  \\
    $E_c$ (MeV) & 1.3$\times10^3$ & 350 & 1 \\
    $\beta$ & 1.68 & 1.92 & 2  \\
    $\eta_\mathrm{em}$ &  & 1.8 &  \\
    \midrule
    \multicolumn{4}{c}{3C279} \\
    $r$ (pc) & 47.9 & 0.016 &   \\
    $B$ (G) & 0.0063 & 2.85 & \\
    $n_e$ (cm$^{-3}$) & 5 & 5.25$\times$10$^4$ &  \\
    $E_c$ (MeV) & 1.6$\times10^3$ & 380 & 1.3 \\
    $\beta$ & 1.75 & 2.05 & 2  \\
    $\eta_\mathrm{em}$ &  & 2.5 &  \\
    \bottomrule\bottomrule
  \end{tabular}
  \label{tab:zones}
\end{table}

 In order to check the self-consistency of our jet and background field models, we calculate steady-state and flaring SEDs for each of our sources with the \texttt{agnpy} \texttt{PYTHON} package\footnote{\url{https://doi.org/10.5281/zenodo.4687123}}, and compare them to broadband observations. We follow the same method as Ref. \cite{Davies_2022}. This is not supposed to be a detailed SED-modeling of our sources (indeed, the spectral parameters and magnetic field strength will be left free in the actual fits), but rather a check that our overall source models are reasonable. To calculate these SEDs, we line up spherical plasma blobs down the jet, each with a field strength (within $\sigma_B$, the errors derived from Ref. \cite{Zamaninasab_2014}), electron density, and bulk Lorentz factor taken from our global jet models (see Table \ref{tab:jet_props}). Every blob contains a population of electrons with a power-law distribution function in energy, up to a cutoff: $N_e(E)=\kappa E^{-\beta}\exp(-E/E_{c})$. The synchrotron emission from all these blobs can be calculated. We can also accelerate electrons (by increasing $E_c$ and adjusting $\beta$) within individual blobs and calculate their synchrotron and inverse-Compton emission using our field models\footnote{With a ring torus model, as implemented in \texttt{agnpy}, placed at the center of our tori, as opposed to our elliptical cross section model. All other field models are the same.} to simulate localized gamma-ray emission regions. For the steady-state emission, we accelerate electrons in the blob at $r_\mathrm{ss}=r_\mathrm{tr}$, as expected from the PC framework, where acceleration is due to a permanent large-scale feature of the jet (e.g., a standing shock, though the detailed acceleration mechanism is not modeled here or in the PC framework). For the flare emission regions, we use a blob located within the jet at $r_\mathrm{em}$, with a radius $R_\mathrm{em}=R/\eta_\mathrm{em}$, where $R$ is the jet width and $\eta_\mathrm{em}>1$. This roughly simulates, e.g., a reconnection or magnetoluminescence event within the highly-magnetized region of the jet, as opposed to a large-scale change in jet structure for the flares, which is disfavoured because of the small flaring timescales.
 Table \ref{tab:zones} shows the parameters used for the various blobs. Figure \ref{fig:model_seds} shows our model SEDs; they are largely consistent with observations in both the flaring and steady-state cases, so we can have confidence in our overall jet and field models.
 This process also enables us to calculate the synchrotron fields within our jets, using the same method as Ref. \cite{Davies_2022}. Each point $r$ will see an isotropic distribution of synchrotron photons from the surrounding blob and an anisotropic one from all the other blobs in the jet. Dispersion off the synchrotron field within the jet is always subdominant, however, and so we do not recalculate it every time $B_0$ changes in the fits.
\section{Systematics}
\label{sec:appendix_sys}
There are also systematic uncertainties associated with the \textit{Fermi} instrument response function, which, as pointed out in Ref. \cite{CTA_Gpropa_2021}, could be important for ALP searches because of the spectral resolution they require. In particular, we are concerned with uncertainties associated with the energy dispersion and reconstruction. Energy dispersion is included in the analysis within \texttt{fermipy}, with different detector response matrices (DRMs) associated with each \texttt{EDISP} class (see Sec. \ref{sec:data}). There are, however, slight uncertainties in these DRMs. In particular, there could be an additional shift and smearing in the reconstructed energy and energy dispersion. These uncertainties can be included with a new expression for the expected counts \cite{CTA_Gpropa_2021}:
\begin{equation}\label{eq:mu_sys}
\begin{split}
    \mu(m_a,g_{a\gamma},E) =& \frac{1}{\mathcal{N}} \int_0^\infty dE' \exp\left(-\frac{(E-E')^2}{2(\delta E)^2}\right) \\ &\times\mu((1-s)E',\mathbf{\theta})\\ &\times P_{\gamma\gamma}(m_a,g_{a\gamma},\mathbf{B},(1-s)E')
\end{split}
\end{equation}
where the parameters $s$ and $\delta$ deal with an energy shift and smear respectively, and
\begin{equation}\label{eq:N_sys}
    \mathcal{N} = \int_0^\infty dE' \exp\left(-\frac{(E-E')^2}{2(\delta E)^2}\right).
\end{equation}
Unfortunately, having to perform this integral at every fine energy that we calculate $P_{\gamma\gamma}$ at greatly increases the computation time, and so it is not feasible to perform the whole analysis in this way---especially as we leave $B_0$ free in the fitting.\par
These errors are expected to be $\lesssim$ a few percent for \textit{Fermi}\footnote{See \href{https://fermi.gsfc.nasa.gov/ssc/data/analysis/LAT_caveats.html}{https://fermi.gsfc.nasa.gov/ssc/data/analysis/LAT\_caveats.html} as accessed on Oct 5, 2022}. Therefore, to test the possible effects of $s$ and $\delta$ on our final results, we perform a single analysis of 3C454.3 using $s=\delta=0.04$.

\begin{figure}
  \centering
    \includegraphics[width=0.48\textwidth]{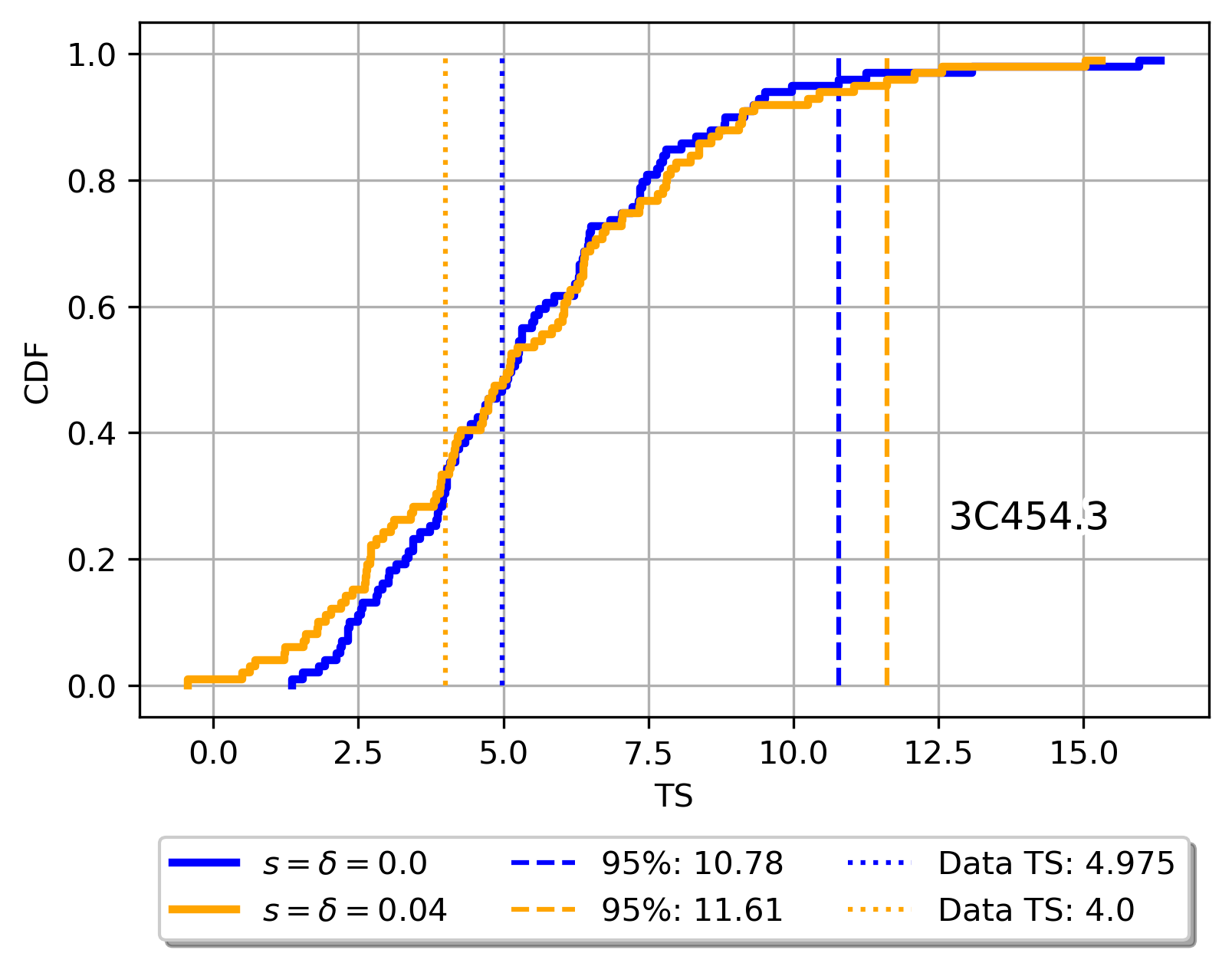}
  \caption{3C454.3 $\mathrm{TS}$ distribution with (orange) and without (blue) additional systematics ($s=\delta=0.04$). Vertical dashed lines show the 95\% thresholds; vertical dotted lines show the $\mathrm{TS}$ value of the data.}
  \label{fig:sys_TS}
\end{figure}

Changes in the systematics would also likely affect the $\lambda_{95}$ thresholds, so the comparison used in Fig. \ref{fig:filled_constraints_alpha}, with the same thresholds as the regular analysis, is less useful in this case.
Nevertheless, we can compare the $\mathrm{TS}$ distributions for the two cases (with and without additional systematics) because the smooth no-ALP spectrum should be unaffected. Figure \ref{fig:sys_TS} shows $\mathrm{TS}$ distributions, for the $s=\delta=0$ (blue) and $s=\delta=0.04$ (orange) cases, for 100 simulated data sets that do not include an injected ALP signal. The vertical dashed lines show the 95\% thresholds, and the vertical dotted lines show the $\mathrm{TS}$ values of the data.
As can be seen, the $\mathrm{TS}$ distributions are very similar in the two cases. An extra shift and (particularly) a smear in the energy reconstruction reduces our ability to distinguish the ALP and no-ALP cases, both slightly reducing the $\mathrm{TS}$ value of the data and slightly increasing the 95\% threshold. Therefore, we would expect the additional systematics to slightly shrink our exclusion regions (in \cite{Fermi_2016}, they show that including similar systematic errors would reduce their exclusion region by $\sim$6\%). This is to be expected, as the Gaussian ($\delta$) term effectively smooths out the oscillations. Nevertheless, even in this case, our observations would likely still constrain ALP parameter space previously unprobed by gamma-ray searches---though more computing power would be required to perform the full analysis.
\end{document}